\documentclass[reprint]{revtex4-1}
\pdfoutput=1
\usepackage[utf8]{inputenc}
\usepackage[T1]{fontenc}
\usepackage{mathtools}
\usepackage{amsfonts}
\usepackage{amssymb}
\usepackage{graphicx}
\usepackage{siunitx}
\usepackage{lmodern}
\usepackage{physics}
\usepackage{microtype}
\usepackage{hyperref}
\usepackage{epstopdf}
\hypersetup{
    linkcolor=blue,
    citecolor=blue,
    urlcolor=blue,
    colorlinks=true,
    pdfauthor={A. A. Golovanov, I. Yu. Kostyukov, L. Reichwein, J. Thomas, A. Pukhov},
    pdftitle={Excitation of strongly nonlinear plasma wakefield by electron bunches}
}
\newcommand{\rb}{r_{\textup{b}}}
\newcommand{\bunch}{{\textup{B}}}
\newcommand{\Rb}{R_{\textup{b}}}
\newcommand{\eff}{{\textup{eff}}}
\newcommand{\submax}{{\textup{max}}}
\newcommand{\submin}{{\textup{min}}}
\newcommand{\plasm}{{\textup{p}}}
\newcommand{\electron}{{\textup{e}}}

\newcommand{\deceleration}{{\textup{dec}}}
\newcommand{\critical}{{\textup{cr}}}
\newcommand{\laser}{{\textup{L}}}
\newcommand{\opt}{{\textup{opt}}}

\begin{document}

\title{Excitation of strongly nonlinear plasma wakefield by electron bunches}
\author{A.\,A. Golovanov}
\author{I.\,Yu. Kostyukov}
\affiliation{Institute of Applied Physics RAS, 603950 Nizhny Novgorod, Russia}
\author{L. Reichwein}
\author{J. Thomas}
\author{A. Pukhov}
\affiliation{Institut für Theoretische Physik I, Heinrich-Heine-Universität Düsseldorf, 40225 Düsseldorf, Germany}

\begin{abstract}
    We propose a new method for analytical self-consistent description of the excitation of a strongly nonlinear wakefield (a bubble) excited by an electron bunch.
    This method makes it possible to calculate the shape of the bubble and the distribution of the electric field in it based only on the properties of the driver, without relying on any additional parameters.
    The analytical results are verified by particle-in-cell simulations and show good correspondence.
    A complete analytical solution for cylindrical drivers and scaling laws for the properties of the bubble and other plasma accelerator parameters depending on the bunch charge and length are derived.
\end{abstract}
\maketitle

\section*{Introduction}

Plasma-based acceleration methods are promising for achieving extremely high energies of charged particles at comparatively short acceleration distances \cite{Esarey_2009_RMP_81_1229, Kostyukov_2015_UFN_58_81}.
They rely on the use of the large longitudinal electric field of the plasma wake.
Such a wake is generated behind a driver propagating in a plasma.
There are two large groups of plasma acceleration methods depending on the type of the driver: laser--wakefield acceleration (LWFA) in which a short laser pulse acts as the driver \cite{Tajima_1979_PRL_43_267}, and plasma--wakefield acceleration (PWFA) in which a charged particle bunch is used to drive the wake \cite{Chen_1985_PRL_54_693, Rosenzweig_1988_PRL_61_98}.
The phase velocity of the wake roughly corresponds to the velocity of the driver propagation, which is usually close to the speed of light.
This allows charged particles to stay in the accelerating phase of the plasma wake for a long period of time and accelerate to high energies.

The physics of interaction of the driver with the plasma is fairly similar in both cases.
A laser driver pushes away plasma electrons from the axis of its propagation due to the action of the averaged ponderomotive force \cite{Quesnel_1998_PRE_58_3719}, while the electron driver does the same with the Coulomb force \cite{Fainberg_1968_PhysUsp_10_750}.
At this time scale, the motion of ions can usually be neglected.
The interaction regime depends on the properties of the driver.
For sufficiently weak drivers, the driven wake is quasi-linear.
However, modern laser technologies based on the chirped pulse amplification technique \cite{Strickland1985compression} make it possible to generate extremely intense femtosecond pulses.
Such pulses, if focussed, interact with plasma in the strongly nonlinear regime, also called the ``bubble'' or ``blowout'' regime.
In this case, the laser driver completely expels electrons from the axis of its propagation, leading to the formation of a spherical cavity devoid of plasma electrons behind it \cite{Pukhov2002Bubble}.
Such a bubble possesses unique properties as an accelerating structure.
Its transverse fields acting on the accelerated electrons creates a focussing transverse force, preventing electrons from leaving the cavity.
And its longitudinal electric field is accelerating in the rear part of the bubble and does not depend on the transverse coordinate, providing more uniform acceleration and reducing the energy spread.
The current record energy achieved in laser--wakefield acceleration in the bubble regime is approximately \SI{8}{GeV} at the acceleration length of just \SI{20}{cm} \cite{Gonsalves_2019_PRL_122_084801}.

A similar strongly nonlinear wake can be achieved with electron drivers when their density is large enough ($n_\bunch \gg n_\plasm$, where $n_\plasm$ is the plasma density), and the size is small ($\sigma_\perp \ll \sigma_\parallel$, $\sigma_\parallel \lesssim \lambda_\plasm$) \cite{Rosenzweig_1991_PRA_44_R6189}.
PWFA offers some significant advantages over LWFA, e.g. much higher phase velocity and the corresponding acceleration length, so its use in the blowout regime is still of great interest despite being comparatively less studied.
In experiments, sufficiently short and high-current bunches can be achieved either by compressing bunches from linear accelerators \cite{Aschikhin_2016_NIMA_806_175} or from a LWFA accelerator, e.g. in hybrid multi-stage LWFA--PWFA accelerators \cite{Martinez_2019_PTRA_377_20180175} or in betatron $\gamma$-ray sources based on the use of bunches from a LWFA stage \cite{Ferri_2018_PRL_120_254802}.
Particles other then electrons can also be used for PWFA.
Impressive results were demonstrated in the AWAKE experiment for a proton driver \cite{Gschwendtner_2019_PTRSA_377_20180418}.
However, obtaining sufficiently short proton bunches to reach the strongly nonlinear regime imposes some challenges \cite{Adli_2016_RAST_09_85} and has not been experimentally demonstrated yet.

Due to the complex nature of the processes and their extremely small temporal and spatial scales which makes experimental diagnostics complicated, experiments on plasma acceleration are usually preceded and supplemented by theoretical studies.
One of the main methods of theoretical analysis are particle-in-cell (PIC) simulations \cite{Birdsall_2004, Pukhov2016PIC}.
Based on solving the Maxwell's equations for the electromagnetic field and the relativistic equations of motion for plasma macroparticles (large clusters of elementary particles), they serve as a tool for ``numerical experiments'' capable of accurately describing all the phenomena accompanying the interaction.
However, full-scale 3D PIC simulations are numerically expensive and do not explain the underlying processses from the theoretical standpoint, so simpler models capable of providing analytical results are also of great interest.

Full theoretical description exists for a quasilinear wake both for laser and particle drivers \cite{Gorbunov_1987_JETP_66_290, Chen_1987_IEEETPS_15_218}. 
However, self-consistent theoretical description of the bubble regime is difficult due to the nonlinear nature of the interaction.
Various blowout regimes of PWFA were described based on numerical simulations \cite{Lotov_2004_PRL_69_046405}.
Early phenomenological models of the bubble regime assumed the ideally spherical shape of the bubble \cite{Kostyukov_2004_PoP_11_115256}.
The use of motion equations for electrons made it possible to describe the shape of the bubble \cite{Lu_2006_PRL_96_165002, Golovanov_2016_QE_46_295} as well as the influence of bunches \cite{Tzoufras_2009_PoP_16_056705}.
This model was later generalized to plasmas with transverse profile \cite{Thomas_2016_PoP_23_053108, Golovanov_2016_PoP_23_093114}.
However, most of these models required the use of some external parameters, preventing the self-consistent solution.

In the current paper, we propose an improvement to the existing models for the case of an electron driver, allowing us to solve the equation for the boundary of the bubble self-consistently.
The solution is derived solely from the properties of the driver and witness bunches.

This paper is structured as follows.
Section~\ref{sec:bubbleEquation} provides the overview of the equation describing the boundary of the bubble.
Section~\ref{sec:energy} describes the first integral of this equation and based on this integral presents an argument for the possibility of starting its solution from the axis.
Section~\ref{sec:excitation} describes the method for solving this equation numerically by overcoming the singularity at the axis.
Section~\ref{sec:analytical} provides a fully analytical result for a cylindrical driver.
In Section~\ref{sec:scaling}, based on the obtained analytical results, scaling laws of the bubble regime of PWFA are derived.

Throughout the paper, the following unitless values are used for simplicity.
Time is normalized to $\omega_\plasm^{-1}$, spatial coordinates to $k_\plasm^{-1} = c / \omega_\plasm$, number densities to $n_\plasm$, charges to $e$, velocities to $c$, the electric and magnetic fields to $m c \omega_\plasm / e$.
Here, $c$ is the speed of light, $\omega_\plasm = (4 \pi e^2 n_\plasm / m)^{1/2}$ is the electron plasma frequency, $n_\plasm$ is the unperturbed electron number density in the plasma, $m$ is the electron mass, and $e > 0$ is the elementary charge.

\section{Problem formulation}
\label{sec:bubbleEquation}

We assume that a wake is generated by an electron driver propagating along the $z$-direction.
For the description of the plasma wake, the evolution of the driver can often be neglected,  and the quasistatic approximation \cite{Sprangle_1990_PRL_64_2011} can be used.
Under this approximation, all values in the wake depend on the co-moving coordinate $\xi = t - z$.
We also assume the cylindrical symmetry of the wake, which leaves the distance to the $z$-axis, $r$, as the second available coordinate.

\begin{figure}[tb!]
    \centering
    \includegraphics[width=\linewidth]{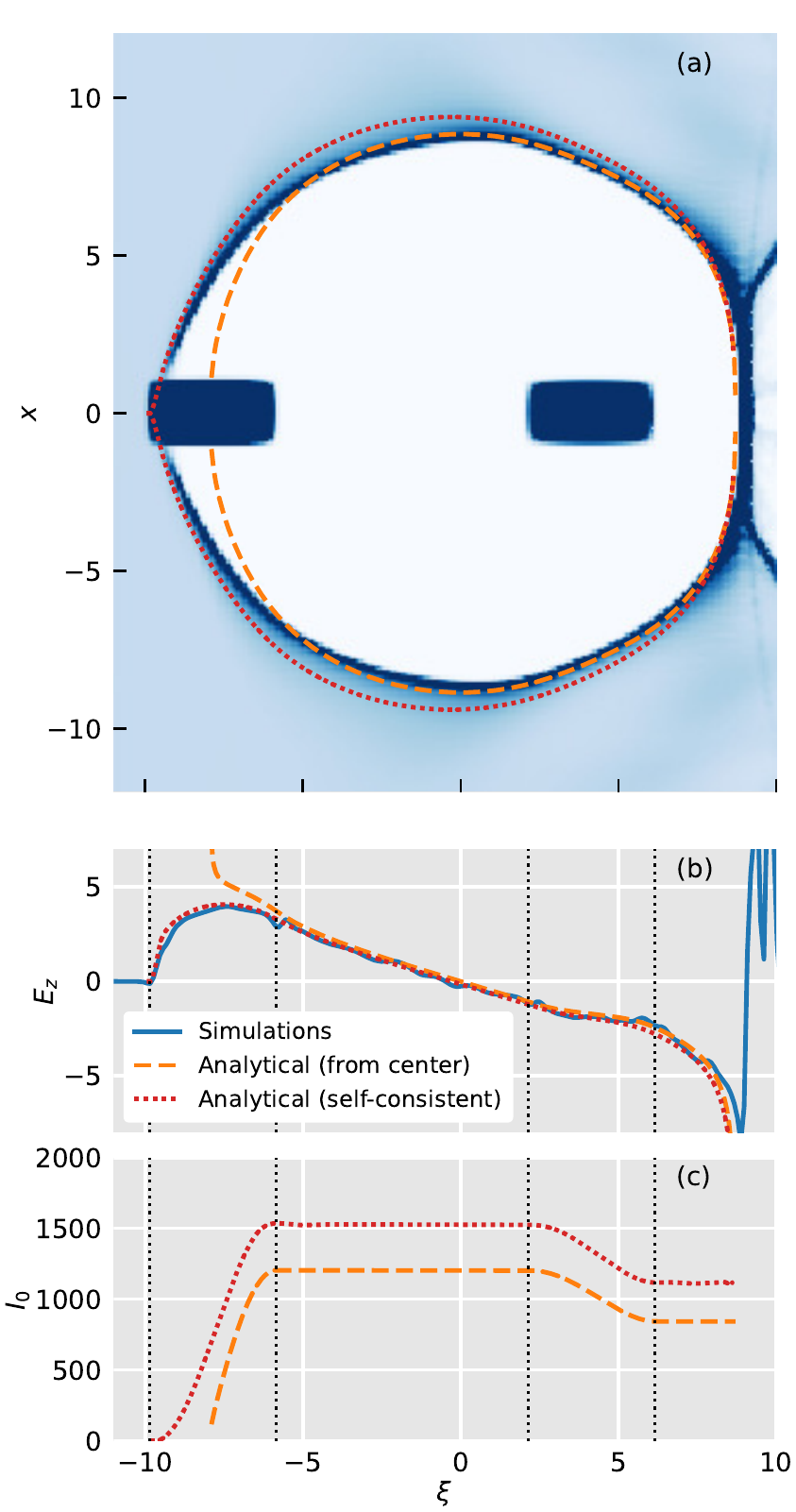}
    \caption{(a) The electron number density distribution in a bubble generated by an electron driver in PIC simulations (see Appendix~\ref{appendix:simulations}) in the $zx$-plane.
    An additional witness bunch is placed inside the bubble.
    Both the driver and the witness propagate to the left.
    (b) The corresponding longitudinal electric field on the axis ($x = 0$) of the bubble.
    (c) The corresponding value of the integral $I_0$ defined by Eq.~\eqref{eq:I0Definition}.
    The dashed lines show the analytical solution to Eq.~\eqref{eq:bubbleEquationUniform} integrated with initial conditions \eqref{eq:bubbleEqInitialConditions}; the dotted lines show the self-consistent analytical solution according to Eq.~\eqref{eq:psiEquation} with initial conditions \eqref{eq:psiEquationInitialConditions}.
    The value $\xi = 0$ corresponds to the cross-section where the bubble reaches its maximum transverse size.
    The vertical lines in (a) and (b) show the boundaries of the bunches.}
    \label{fig:bubbleExample}
\end{figure}

In the strongly nonlinear regime of plasma wakefield, the driver generates a bubble devoid of plasma electrons behind it (see Fig.~\ref{fig:bubbleExample}a).
On the boundary of the bubble, a thin electron sheath is formed; it screens the bubble from the surrounding unperturbed plasma.
The boundary between the inner part of the bubble and the sheath is very abrupt, and it can be described by a scalar function of the longitudinal coordinate, $\rb(\xi)$, satisfying the equation \cite{Thomas_2016_PoP_23_053108, Golovanov_2016_QE_46_295}
\begin{equation}
    A(\rb) \dv[2]{\rb}{\xi} + B(\rb) \qty(\dv{\rb}{\xi})^2 + C(\rb) = \lambda(\xi, \rb).
    \label{eq:bubbleEquationGeneral}
\end{equation}
The term on the right-hand side,
\begin{equation}
    \lambda(\xi, r) = -\int_0^{r} {\rho_\bunch(\xi, r') r' \dd{r'}},
    \label{eq:lambdaDef}
\end{equation}
describes the influence of bunches (both driver and witness) inside the bubble on its boundary.
Coefficients $A$, $B$, $C$ generally depend on the transverse plasma profile \cite{Thomas_2016_PoP_23_053108} and the properties of the electron sheath on the boundary of the bubble \cite{Golovanov_2016_QE_46_295}.
In this paper, we will limit ourselves to uniform plasma without any transverse profiling.
In this case, the coefficients are
\begin{align}
    &A(\rb) = \rb \left(1 + \frac{\rb^2}{4} + \rb^2 \beta + \frac{\rb^3}{4} \dv{\beta}{\rb} \right),\\
    &B(\rb) = \frac{\rb^2}{2} \left(1 + 3 \beta + 3 \rb \dv{\beta}{\rb} + \frac{\rb^2}{2} \dv[2]{\beta}{\rb} \right),\\
    &C(\rb) = \frac{\rb^2}{4} \frac{1 + (1+\rb^2\beta/2)^2}{(1 + \rb^2 \beta/2)^2},
\end{align}
where $\beta(\rb)$ is a function characterizing the electron sheath on the boundary of the bubble.
This function depends on the width of the electron sheath $\Delta$ and its transverse profile (see Ref.~\cite{Golovanov_2016_QE_46_295} for more details).
In most cases, $\beta(\rb)$ can be approximated by a function $\beta(\rb) \approx \Delta / \rb$.

If the width $\Delta$ of the electron sheath on the boundary of the bubble satisfies two additional conditions: $\Delta \ll \rb$ and $\Delta \gg \rb^{-1}$, then $\beta \ll 1$ and $\beta \rb^2 \gg 1$.
This approximation is called the relativistic approximation \cite{Lu_2006_PRL_96_165002, Golovanov_2016_QE_46_295}, as it corresponds to electrons in the electron sheath being ultra-relativistic.
If we use this approximation and neglect all insignificant terms in coefficients $A$--$C$, Eq.~\eqref{eq:bubbleEquationGeneral} is reduced to \cite{Lu_2006_PRL_96_165002}
\begin{equation}
    \rb \dv[2]{\rb}{\xi} + 2 \qty(\dv{\rb}{\xi})^2 + 1 = 2 \varkappa(\xi, \rb).
    \label{eq:bubbleEquationUniform}
\end{equation}
To simplify further calculations, we have introduced a new function
\begin{equation}
    \varkappa(\xi, r) = \frac{2}{r^2} \lambda(\xi, r) = - \frac{2}{r^2} \int_0^r {\rho_\bunch(\xi, r') r' \dd{r'}}.
    \label{eq:kappa}
\end{equation}
From this definition, it is obvious that 
\begin{equation}
    \lim_{r \to 0} \varkappa(\xi, r) = - \rho_\bunch(\xi, 0).
\end{equation}
For electron bunches with negative charge density, we have $\varkappa > 0$.
So $\varkappa(\xi, r)$ corresponds to the average absolute charge density of the bunch inside the circle with the radius $r$ in some cross-section $\xi$.

The longidutinal electric field $E_z$ responsible for the acceleration or deceleration of particles does not depend on the radial coordinate \cite{Lu_2006_PRL_96_165002, Thomas_2016_PoP_23_053108}.
In a bubble described by Eq.~\eqref{eq:bubbleEquationUniform}, it can be calculated as 
\begin{equation}
    E_z(\xi) = \frac{\rb}{2} \dv{\rb}{\xi}.
    \label{eq:Ez}
\end{equation}
Transverse fields can also be calculated from the shape of the bubble boundary \cite{Yi_2013_PoP_20_013108, Golovanov_2017_PoP_24_103104}, but this is beyond the scope of the current paper.

As we have already stated, Eq.~\eqref{eq:bubbleEquationUniform} is written in the relativistic approximation when $\Delta \ll \rb$, $\Delta \gg \rb^{-1}$, which automatically implies that $\rb \gg 1$.
Therefore, the equation should be invalid in the area where $\rb$ is close to zero.
That is why it is common to solve this equation starting from the center of the bubble, i.e. the point where the bubble reaches its maximum transverse size $\Rb$ \cite{Tzoufras_2009_PoP_16_056705, Thomas_2016_PoP_23_053108, Golovanov_2016_PoP_23_093114}.
The initial conditions in this case are
\begin{equation}
    \rb(\xi_0) = \Rb, \quad \dv{\rb}{\xi}\Bigg|_{\xi = \xi_0} = 0.
    \label{eq:bubbleEqInitialConditions}
\end{equation}

The value of $\Rb$ as well as the point $\xi_0$ where $\rb$ reaches its maximum are generally unknown and have to be set manually.
If we choose those two values according to the results of PIC simulations in Fig.~\ref{fig:bubbleExample}, the solution to Eq.~\eqref{eq:bubbleEquationUniform} with initial conditions \eqref{eq:bubbleEqInitialConditions} as well matches the result of simulations very well (see the dashed lines in Fig.~\ref{fig:bubbleExample}a).
The same is true for the longitudinal electric field (dashed lines in Fig.~\ref{fig:bubbleExample}b).
The discrepancy arises in the frontal and rear parts of the bubble where $\rb$ approaches $0$, and the assumption $\rb \gg 1$ fails.
This good correspondence was demonstrated in many previous works \cite{Lu_2006_PRL_96_165002, Tzoufras_2009_PoP_16_056705, Thomas_2016_PoP_23_053108, Golovanov_2016_PoP_23_093114}, which justifies the use of Eq.~\eqref{eq:bubbleEquationUniform} written under the relativistic approximation in theoretical studies.
This approximation generally fails to describe the bubble only when the its maximum size $\Rb \sim 1$, and the condition $\rb \gg 1$ is not met anywhere.

However, the main problem of this approach is the necessity to rely on the results of PIC simulations to find the center point $\xi_0$ and the transverse size $\Rb$ of the bubble.
Without this, we do not know the initial conditions Eq.~\eqref{eq:bubbleEqInitialConditions}.
In this sense, it is not a self-consistent solution, as additional manually set parameters are required to find a solution for a certain driver.
For practical purposes, it would be much more useful to find a self-consistent solution, i.e. a solution relying only on the properties of the driver.
As Eq.~\eqref{eq:bubbleEquationUniform} is invalid around small values $\rb$, it is seemingly impossible to use this equation to find the solution in the self-consistent manner, which basically requires starting the solution from $\rb = 0$.
Next, we present an argument why using this equation at $\rb = 0$ can still be justified.

\section{Energetic properties of the bubble}
\label{sec:energy}

In the regions along the $\xi$-axis where there are no bunches and therefore $\varkappa(\xi, r) = 0$, Eq.~\eqref{eq:bubbleEquationUniform} has a first integral:
\begin{equation}
    I_0(\xi) = \pi \frac{\rb^4}{16} \left[1 + 2 \qty(\dv{\rb}{\xi})^2 \right] = \pi \frac{\Rb^4}{16} = \text{const},
    \label{eq:I0Definition}
\end{equation}
where $\Rb$ is the maximum size of the bubble.

If $\varkappa \neq 0$, then the value of the integral changes,
\begin{equation}
    \dv{I_0}{\xi} = \frac{\pi}{2} \rb^3 \dv{\rb}{\xi} \varkappa(\xi, \rb).
    \label{eq:I0equation}
\end{equation}
If we integrate this formula and use Eq.~\eqref{eq:Ez} for $E_z$ and the definition of $\varkappa$ \eqref{eq:kappa}, we get
\begin{multline}
    I_0(\xi) - I_0(\xi_0)\\ = - 2 \pi \int_{\xi_0}^\xi \int_0^{\rb} {E_z(\xi') \rho_\bunch(\xi', r') r' \dd{r'} \dd{\xi'}}.
    \label{eq:I_0_integral}
\end{multline}
The integral on the right-hand side is essentially equal to $-\int \vb{j}_\bunch \vb{E} \dd{V}$, which is the power of the energy exchange between the electric field of the bubble and the bunch within the volume between $\xi_0$ and $\xi$.
Therefore, $I_0$ serves as a measure of the energy density in the bubble.
The dashed lines in Fig.~\ref{fig:bubbleExample}c show the value of $I_0$ corresponding to the analytical solution shown in Fig.~\ref{fig:bubbleExample}a,b.
In the areas where there are no electron bunches, the value of $I_0$ remains constant.
For the witness bunch, when the field is accelerating for electrons ($E_z < 0$), $I_0$ decreases with $\xi$ which corresponds to expending the energy of the bubble on acceleration.
On the contrary, when the field is decelerating ($E_z > 0$), $I_0$ increases with $\xi$, corresponding to the driver pumping the bubble.

As the energy density in the bubble is determined by a single parameter, its maximum size $\Rb$, it is also convenient to introduce the local effective size of the bubble
\begin{equation}
    R_\eff(\xi) = 2 \qty(\frac{I_0}{\pi})^{1/4} = \rb \left[1 + 2 \qty(\dv{\rb}{\xi})^2 \right]^{1/4}.
    \label{eq:Reff}
\end{equation}
In this case, we can say that the driver gradually increases the size of the bubble, while the witness reduces it.

The physical meaning of $I_0$ becomes more evident if we use the dimensional values.
In this case, $I_0$ is defined as
\begin{equation}
    I_0 = \frac{m^2 c^5}{4\pi e^2} \pi \frac{(k_\plasm R_\eff)^4}{16} = \frac{m^2 c^5}{64 e^2} (k_\plasm R_\eff)^4, 
    \label{eq:I0Physical}
\end{equation}
and satisfies the equation
\begin{equation}
    I_0(\xi) - I_0(\xi_0) = P = - \int \vb{j} \vb{E} \dd{V}.
\end{equation}
From this definition, we see that $I_0$ is normalized to the units of $m^2 c^5/(4 \pi e^2)$ and has the dimension of \emph{power}.
It is equal to the power corresponding to the energy loss by the driver, and it is also equal to the maximum possible power of acceleration achievable in this bubble (as the witness cannot make $I_0$ negative).
In fact, it can be shown that $I_0$ is also equal to
\begin{align}
    &I_0 = 2\pi \int_0^{\rb} \left[c W - S_z \right] r' \dd{r'}, \\
    &W = \frac{\vb{E}^2 + \vb{B}^2}{8 \pi}, \quad \vb{S} = \frac{c}{4\pi} \vb{E} \cp \vb{B},
\end{align}
where $W$ is the electromagnetic energy density, and $\vb{S}$ is the Poynting vector.
Thus, $I_0$ has a physical meaning of the energy flux along the comoving window over the cross-section of the bubble.
Energetic properties and the energy flux in the blowout regime are studied in detail in Ref.~\cite{Lotov_2004_PRL_69_046405}, and the same quantity $\pi \Rb^4/16$ arises there from the energetic relations alone.

In Fig.~\ref{fig:bubbleExample}c, because the solution (shown with the dashed line) is not self-consistent, $I_0$ never reaches $0$, as $\rb$ collapses to $0$ already inside the bunch.
This obviously contradicts the reality, as the frontal part of the driver should also perform work over the wakefield and contribute to $I_0$.
For a self-consistent solution, we should assume that initially $I_0$ is equal to $0$, corresponding to $\rb = 0$, and the driver begins to ``pump'' the wakefield immediately.
And despite the fact that the resulting solution is seemingly invalid around $\rb < 1$, for sufficiently dense drivers, $\rb$ will quickly reach large enough values, and the rest of the integral in Eq.~\eqref{eq:I_0_integral} will be calculated in the area where $\rb \gg 1$, and our assumptions hold.
Therefore, we might expect that the final value of $I_0$ in such a self-consistent solution will be close to the correct value, despite the fact that we have started from the area where Eq.~\eqref{eq:bubbleEquationUniform} is not applicable.
That this expectation is indeed met will be shown in the following sections.

\section{Equation for bubble excitation}
\label{sec:excitation}

We have presented an argument that Eq.~\eqref{eq:bubbleEquationUniform} should be applicable to the description of the excitation of the bubble.
However, its direct numerical solution from $\rb = 0$ is difficult because, when $\rb = 0$, the second derivative $\dv*[2]{\rb}{\xi}$ diverges.

To circumvent this problem, we make a substitution,
\begin{equation}
    \psi_\xi(\xi) = \frac{\rb^2(\xi)}{4}.
    \label{eq:psiDefinition}
\end{equation}
The quadratic substitution is the only power-law substitution that eliminates the divergence of the second derivative without introducing a zero solution into the equation.
This new quantity is related to the wakefield potential, $\psi = \varphi - A_z$, which in the bubble in uniform plasma can be calculated as
\begin{equation}
    \psi(\xi,r) = \frac{\rb^2(\xi) - r^2}{4}.
\end{equation}
Therefore, $\psi_\xi(\xi)$ is the independent of $r$ component of $\psi$, which explains why we use the additional $1/4$ factor in the substitution.
As expected,
\begin{equation}
    E_z(\xi) = \pdv{\psi}{\xi} = \dv{\psi_\xi}{\xi},
\end{equation}
which gives us the same result as Eq.~\eqref{eq:Ez}.

If we substitute $\rb(\xi)$ with $\psi_\xi(\xi)$ in Eq.~\eqref{eq:bubbleEquationUniform}, we get
\begin{equation}
    \dv[2]{\psi_\xi}{\xi} + \frac{1}{2\psi_\xi} \qty(\dv{\psi_\xi}{\xi})^2 = \varkappa(\xi, 2 \sqrt{\psi_\xi}) - \frac{1}{2}.
    \label{eq:psiEquation}
\end{equation}
In this equation, the second derivative is now finite, so we use it for numerical integration instead of using Eq.~\eqref{eq:bubbleEquationUniform}.
For the problem of excitation, we have to solve this equation with the initial conditions
\begin{equation}
    \psi_\xi(\xi_0) = 0, \quad \dv{\psi_\xi}{\xi}\Bigg|_{\xi_0=0} = 0.
    \label{eq:psiEquationInitialConditions}
\end{equation}
We also demand that the second term satisfies the condition
\begin{equation}
    \frac{1}{2\psi_\xi} \qty(\dv{\psi_\xi}{\xi})^2\Bigg|_{\rb=0} = 0.
\end{equation}
As $\psi_\xi$, by definition, is a strictly positive quantity, the right-hand side of Eq.~\eqref{eq:psiEquation} has to be initially positive.
This means that the excitation of the bubble can occur only if
\begin{equation}
    \varkappa(\xi, 0) = \abs{\rho_\bunch(\xi, 0)} > \frac{1}{2}.
\end{equation}
This is similar to the necessary condition of reaching the blowout regime found in the literature, $\abs{\rho_\bunch} > 1$ \cite{Rosenzweig_1991_PRA_44_R6189}.
However, the value of $1/2$ should not have any special meaning, as Eq.~\eqref{eq:psiEquation} is not strictly valid around $\psi_\xi = 0$.

In a real electron bunch, the charge density increases gradually from $0$ to its peak value along the bunch.
So we should start solving Eq.~\eqref{eq:psiEquation} from a point $\xi_0$ where $\abs{\rho_\bunch(\xi_0, 0)} = 1/2$.
The initial behavior around this point can be calculated analytically to verify our assumptions about the properties of the solution are correct (see Appendix~\ref{appendix:analyticalExcitation}).
However, in order to find the entire solution, we generally have to rely on numerical integration.

After we have calculated $\psi_\xi(\xi)$, we can restore all other quantities,
\begin{align}
    &\rb = 2 \sqrt{\psi_\xi}, \quad \dv{\rb}{\xi} = \frac{1}{\sqrt{\psi_\xi}} \dv{\psi_\xi}{\xi}, \quad E_z = \dv{\psi_\xi}{\xi},\\
    &I_0(\xi) = \pi \psi_\xi^2 \left[1 + \frac{2}{\psi_\xi} \qty(\dv{\psi_\xi}{\xi})^2 \right].
\end{align}

The comparison between the results of PIC simulations, the numerical integration of Eq.~\eqref{eq:psiEquation} with initial conditions \eqref{eq:psiEquationInitialConditions} and numerical integration of Eq.~\eqref{eq:bubbleEquationUniform} from the center of the bubble is shown in Fig.~\ref{fig:bubbleExample}a--c.
The details on PIC simulation parameters can be found in Appendix~\ref{appendix:simulations}.
In the self-consistent solution (dotted line in Fig.~\ref{fig:bubbleExample}a--c), $I_0$ is calculated from the value of $0$ at the front, and the peak value becomes slightly larger than in the solution calculated from the center of the bubble (dashed lines).
This also results in a slightly larger than expected size of the bubble.
However, the longitudinal electric field (Fig.~\ref{fig:bubbleExample}b) shows much better correspondence in the self-consistent case rather than the integration from the center of the bubble, where the electric field diverges inside the bunch.
This happens even though Eq.~\eqref{eq:psiEquation} is supposed to be inaccurate for small values of $\psi_\xi$ (or $\rb$).
Another advantage of the new approach, as already stated earlier, is its self-consistent nature.
To integrate Eq.~\eqref{eq:psiEquation}, we only use the properties of the driver, and do not use any external parameters like the size of the bubble $\Rb$ or the position of its center.
These quantities are found in a self-consistent manner.

\begin{figure*}[tb]
    \includegraphics[width=\linewidth]{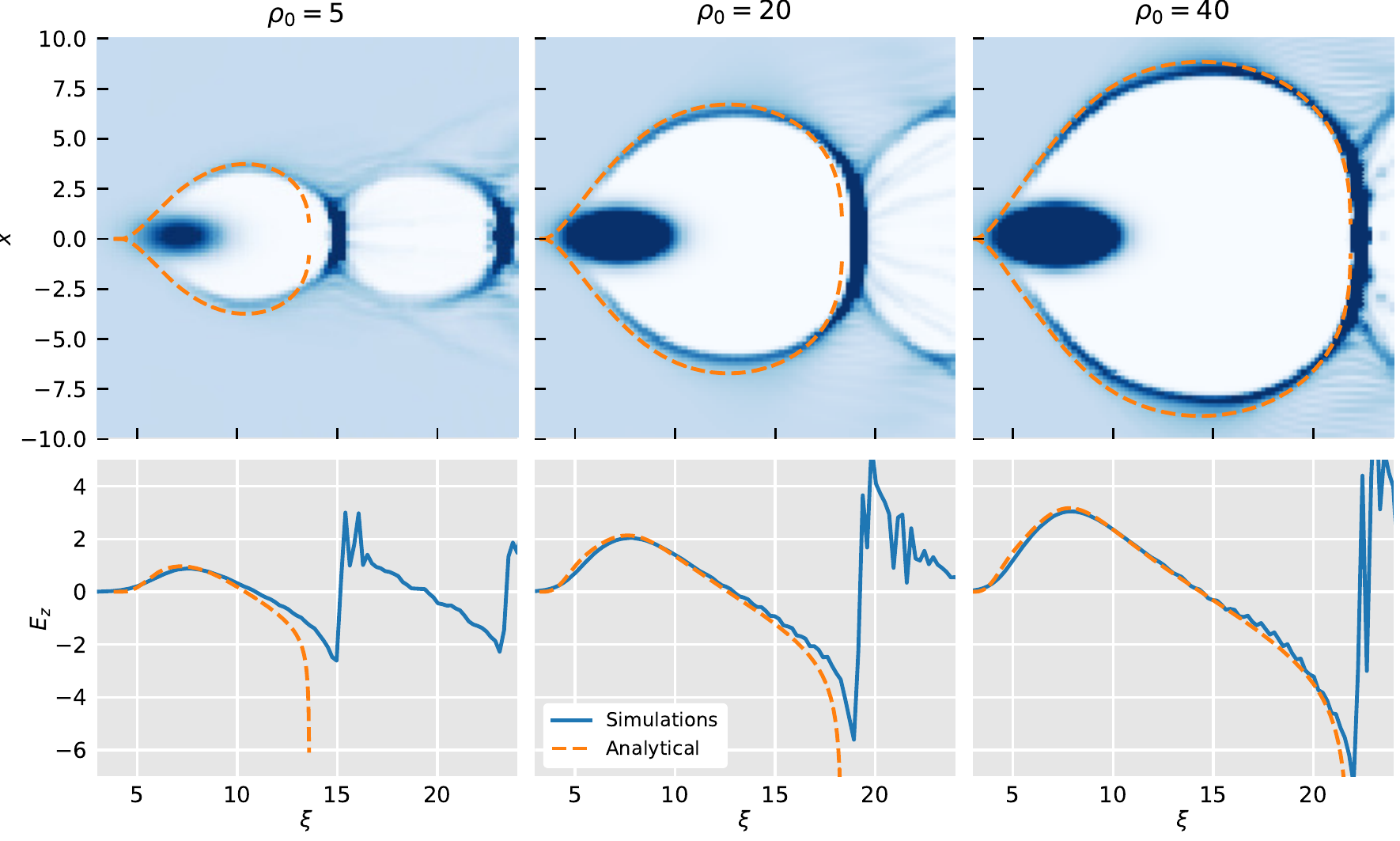}
    \caption{The electron number density $n_\electron$ in the $zx$-plane (top) and the longitudinal electric field $E_z$ on the axis $x = 0$ in bubbles generated by an electron driver with the maximum charge densities $\rho_0$ of $5$, $20$, and $40$ as observed in PIC simulations (see Appendix~\ref{appendix:simulations}).
    The dashed lines show the solutions according to Eq.~\eqref{eq:psiEquation} with the initial conditions \eqref{eq:psiEquationInitialConditions}.}
    \label{fig:bubbles-gaussian}
\end{figure*}

Another series of comparisons between the analytical predictions from Eq.~\eqref{eq:psiEquation} and 3D PIC simulations for different drivers is shown in Fig.~\ref{fig:bubbles-gaussian}.
For the weakest driver ($\rho_0 = 5$), when the maximum size of the bubble is small, the correspondence between the analytical results and the simulations is not very good because the assumption that $\rb \gg 1$ is generally invalid everywhere.
In this case, the bubble in PIC simulations is elongated compared to the nearly spherical bubble predicted by the simplified theoretical model.
However, as the density of the bunch grows, the correspondence becomes more and more accurate.
So, for sufficiently large bubbles, the proposed theory can be used to describe the properties of the excited bubble and the structure of the accelerated field in a self-consistent manner based only on the properties of the driver and without the use of any external parameters.

\section{Solution for a cylindrical driver}
\label{sec:analytical}

In general, Eq.~\eqref{eq:psiEquation} can be solved numerically for an arbitrary driver.
However, it might be helpful to have an analytical solution for some special case to better understand the physics of bubble excitation and the limitations of the suggested model.

To do so, we consider a cylindrical bunch with the longitudinal size of $\xi_\bunch$ and the transverse size of $r_\bunch$,
\begin{equation}
    \rho_\bunch(\xi, r) = \begin{cases}
        - \rho_0, & r < r_\bunch \text{ and } 0 \leq \xi \leq \xi_\bunch,\\
        0, & \text{elsewhere.}
    \end{cases}
\end{equation}
In this case,
\begin{equation}
    \varkappa(\xi, r) = \begin{dcases}
        \rho_0, & r < r_\bunch \text{ and } 0 \leq \xi \leq \xi_\bunch,\\
        \rho_0 \frac{r_\bunch^2}{r^2}, & r \geq r_\bunch \text{ and } 0 \leq \xi \leq \xi_\bunch,\\
        0, & \text{elsewhere.}
    \end{dcases}
\end{equation}
To actually have a solution to Eq.~\eqref{eq:psiEquation}, we assume that $\rho_0 > 1/2$.

First, we find the solution inside the bunch, where $\varkappa = \rho_0$.
This case is different from the excitation process described in Section~\ref{sec:excitation} and Appendix~\ref{appendix:analyticalExcitation}, as there is no gradual growth of the density, it abruptly jumps to the value of $\rho_0$.
Thus, we can no longer use the same assumptions.
For instance, the assumption that the second term in Eq.~\eqref{eq:psiEquation} is initially equal to zero is invalid in this case.
However, the analytical solution is easy to guess from Eq.~\eqref{eq:bubbleEquationUniform} directly,
\begin{equation}
    \rb(\xi) = \sqrt{\tilde{\rho_0}} \xi, \quad \dv{\rb}{\xi} = \sqrt{\tilde{\rho}_0}, \quad E_z = \frac{\tilde{\rho}_0 \xi}{2},
    \label{eq:analytical:insidebunch}
\end{equation}
where $\tilde\rho_0 = \rho_0 - 1/2$.
Under the assumption that the bunch is narrow and long enough, $r_\bunch < \sqrt{\tilde{\rho}_0} \xi_\bunch$, this solution is valid until the point $\xi_1$ defined by $r_\bunch = \sqrt{\tilde{\rho}_0} \xi_1$.

Next, we have to find the solution between points $\xi_1$ and $\xi_\bunch$, where $\varkappa = \rho_0 r_\bunch^2/r^2$.
To do so, we use Eq.~\eqref{eq:I0equation} for the evolution of $I_0$,
\begin{equation}
    \dv{I_0}{\xi} = \frac{\pi}{2} r_\bunch \rb \dv{\rb}{\xi}.
\end{equation}
As the right-hand side is the full derivative of $\rb^2$, we can easily integrate it.
Taking the initial conditions from the solution inside the bunch \eqref{eq:analytical:insidebunch}, we get
\begin{equation}
    \qty(\dv{\rb}{\xi})^2 = - \frac{1}{2} - \frac{\rho_0 r_\bunch^4}{\rb^4} + \frac{2 \rho_0 r_\bunch^2}{\rb^2}.
\end{equation}
It is convenient to introduce new variables,
\begin{equation}
    u = \frac{\rb}{r_\bunch}, \quad \zeta = \frac{\xi}{r_\bunch}
\end{equation}
to remove $r_\bunch$ from the equation,
\begin{equation}
    \qty(\dv{u}{\zeta})^2 = -\frac{1}{2} - \frac{\rho_0}{u^4} + \frac{2 \rho_0}{u^2}.
\end{equation}
The initial condition for this equation is $u(\zeta_1) = 1$.
We will find the solution only in the area where $\dv*{u}{\zeta} > 0$.
The solution after $u$ reaches its maximum is just a symmetrical continuation, as changing $\zeta \to -\zeta$ does not change the equation.
We also rewrite the equation as
\begin{multline}
    \sqrt{2} u^2 \dv{u}{\zeta} 
    \\ = \sqrt{(2\rho_0 + 2 \sqrt{\rho_0 \tilde{\rho}_0} - u^2)(u^2 - 2 \rho_0 + 2\sqrt{\rho_0 \tilde{\rho}_0})}.
    \label{eq:analytical:uEquation}
\end{multline}
The maximum value of $u$ is reached when the right-hand side turns to zero.
It can be shown that $2\rho_0 - 2 \sqrt{\rho_0 \tilde{\rho}_0} < 1$ for any value of $\rho_0$.
As $u \geq 1$, the second multiplier is thus always positive and cannot be equal to $0$.
Thus the maximum possible value is determined by setting the first multiplier to zero,
\begin{equation}
    u_\submax = \sqrt{2\rho_0 + 2 \sqrt{\rho_0 \tilde{\rho}_0}}.
\end{equation}
If $\rho_0$ is sufficiently large, $\tilde{\rho}_0 \approx \rho_0$, and $u_\submax \approx 4 \sqrt{\rho_0}$.
Thus, the maximum possible radius of a bubble excited by a cylindrical driver with the density of $\rho_0$ and the radius of $r_\bunch$ cannot exceed the value of
\begin{equation}
    R_\submax = 4 r_\bunch \sqrt{\rho_0},
\end{equation}
regardless of the length of such a driver.
This value is proportional to the square root of $\rho_0 r_\bunch^2$, or the total current of the bunch.
As the bubble should be described by the relativistic approximation, we must demand
\begin{equation}
    R_\submax \gg 1, \quad 4\sqrt{\rho_0 r_\bunch^2} \gg 1,
\end{equation}
which is the condition for the total current which limits the applicability of our model.

Next, we could try solving Eq.~\eqref{eq:analytical:uEquation} directly, but we can simplify it using the assumption that $\rho_0 \gg 1/2$, which is usually satisfied for realistic drivers.
In this case, Eq.~\eqref{eq:analytical:uEquation} can be simplified to
\begin{equation}
    \sqrt{2} u \dv{u}{\zeta} = \sqrt{2 \rho_0 + 2 \sqrt{\rho_0 \tilde{\rho}_0} - u^2},
\end{equation}
and its solution, after returning to $\rb$ and $\xi$, is 
\begin{equation}
    \frac{\rb}{r_\bunch} = \sqrt{1  + \sqrt{4\tilde{\rho}_0 + 4 \sqrt{\tilde{\rho_0} \rho_0}} \frac{\xi - \xi_1}{r_\bunch} - \frac{(\xi - \xi_1)^2}{2 r_\bunch^2}}.
    \label{eq:analytical:approximate_rb}
\end{equation}
Here, we have preserved $\tilde{\rho}_0$ in order to have exactly the same value of $u_\submax$ as in the exact solution.

\begin{figure}[tb!]
    \includegraphics[width=\linewidth]{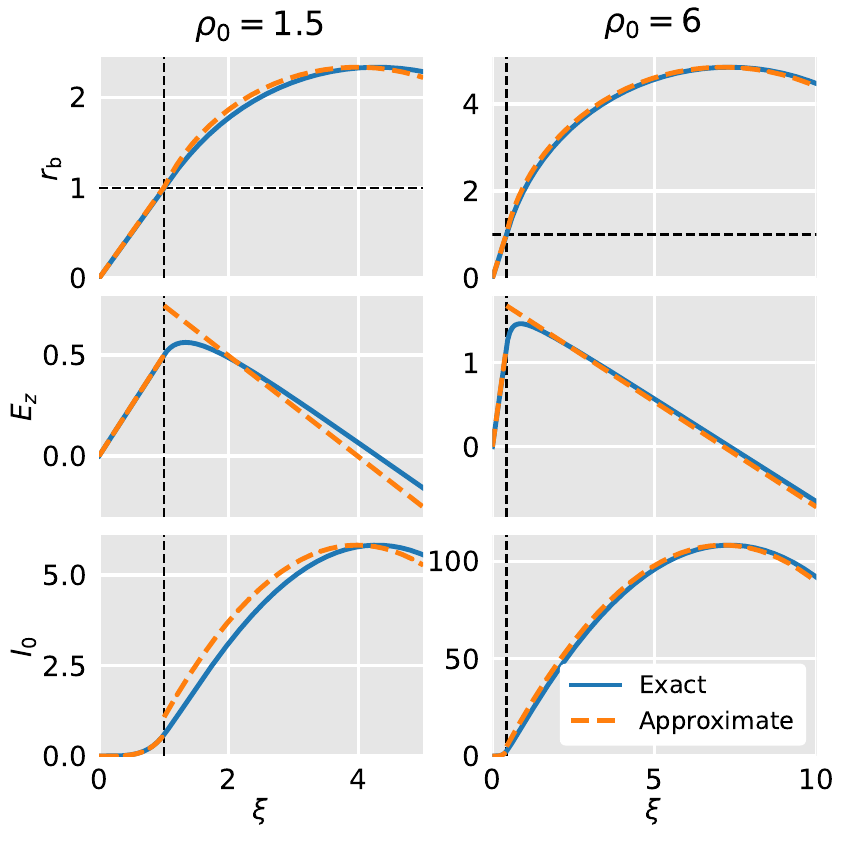}
    \caption{The dependencies of the bubble radius $\rb$, the electric field $E_z$, and the integral $I_0$ on the longitudinal coordinate $\xi$ for a cylindrical driver with $\rho_0$ equal to 1.5 and 6.
    The radius of the driver $r_\bunch = 1$ (shown by the horizontal lines), the length of the driver $\xi_\bunch \to \infty$ is not limited.
    The solid lines show the exact solution according to Eq.~\eqref{eq:analytical:uEquation}, the dashed lines correspond to the approximate solution \eqref{eq:analytical:approximate_rb}.
    Solutions inside the driver ($\rb < 1$) are given by Eq.~\eqref{eq:analytical:insidebunch}.
    The vertical lines correspond to the point $\xi_1$ when $\rb$ reaches the boundary $r_\bunch=1$ of the driver.}
    \label{fig:analytical_solutions}
\end{figure}

The comparison of the exact numerical solution to Eq.~\eqref{eq:analytical:uEquation} and the approximate solution given by Eq.~\eqref{eq:analytical:approximate_rb} is shown in Fig.~\ref{fig:analytical_solutions} for $\rho_0$ equal to $1.5$ and $6$.
The simplification of Eq.~\eqref{eq:analytical:uEquation} leads to a slight discontinuity in $E_z$ and $I_0$ at $\xi = \xi_1$.
However, even for values of $\rho_0$ as low as $1.5$, the difference between the exact and the approximate solutions is pretty small, and it vanishes with the increase of $\rho_0$.
Therefore, in a more realistic case of $\rho_0 \gg 1$, solution $\eqref{eq:analytical:approximate_rb}$ is a good approximation.
By the way, this solution is also correct even if the derivative of $\rb$ switches sign, $\dv*{\rb}{\xi} < 0$.

Solution~\eqref{eq:analytical:approximate_rb} is valid until $\xi$ reaches the value of $\xi_\bunch$.
For $\xi > \xi_\bunch$, where $\varkappa = 0$, the solution can also be found analytically, but we will not be interested in it because the driver no longer contributes to the increase of the bubble's power in this region, and $I_0(\xi > \xi_\bunch) = I_0(\xi_\bunch)$.

Finally, based on the obtained solutions, let us write the final $I_0$ for a cylindrical driver with the density $\rho_0$ and the sizes $\xi_\bunch$ and $r_\bunch$,
\begin{multline}
    I_0 = \pi \frac{r_\bunch^4 (\rho_0 + \sqrt{\rho_0 \tilde{\rho}_0})}{8}
    \\ \times \left[1  + \sqrt{4\tilde{\rho}_0 + 4 \sqrt{\tilde{\rho_0} \rho_0}}
    \frac{\xi_\bunch - \xi_1}{r_\bunch} - \frac{(\xi_\bunch - \xi_1)^2}{2 r_\bunch^2} \right].
\end{multline}
When $\rho_0 \gg 1$, 
\begin{equation}
    I_0 \approx \pi \frac{\rho_0 r_\bunch^2}{8} \left[4 \sqrt{2\rho_0 r_\bunch^2} \xi_\bunch - \xi_\bunch^2 \right].
    \label{eq:analytical:I0_large_rho}
\end{equation}
In this limit, $I_0$ is determined by only two parameters: the length of the driver $\xi_\bunch$ and its total current proportional to $\rho_0 r_\bunch^2$.

We can also replace the total current with the absolute total charge,
\begin{equation}
    Q = \pi \rho_0 r_\bunch^2 \xi_\bunch.
\end{equation}
In this case,
\begin{equation}
    I_0 = Q \left[\sqrt{\frac{Q}{2\pi\xi_\bunch}} - \frac{\xi_\bunch}{8}\right].
    \label{eq:analytical:I0_analytical_Q}
\end{equation}

From this solution we can conclude that, if we maintain the constant value of the bunch charge $Q$ instead of the constant value of the density $\rho$ and vary only its length $\xi_\bunch$, the most optimal bunch is the shortest possible.
At the same time, the maximum possible length of the driver is limited by a condition $I_0 > 0$,
\begin{equation}
    \xi_{\bunch,\submax} = \left(\frac{32 Q}{\pi}\right)^{1/3}.
\end{equation}
A bunch longer than this will not fit into a bubble it generates.
Generally, a length much shorter than this is desirable.
If $\xi_\bunch \ll \xi_{\bunch,\submax}$, then
\begin{equation}
    I_0 \approx \frac{Q^{3/2}}{\sqrt{2\pi \xi_\bunch}}.
\end{equation}

Although the obtained solution is derived for a cylindrical driver, it is expected that the solution for an arbitrary driver will be qualitatively the same, only the constants will be different.

\section{Scaling laws for the electron driver}
\label{sec:scaling}

The obtained analytical results would be more convenient to analyze and use if we return to the dimensional units in order to make the plasma density an explicit parameter.
First of all, from Eq.~\eqref{eq:I0Physical}, we directly see that the normalization units for $I_0$ are independent of the plasma properties,
\begin{equation}
    \frac{m^2 c^5}{4\pi e^2} \approx \SI{6.93e15}{erg/s} = \SI{693}{MW}
\end{equation}
This constant is $8\pi$ times lower then another more widely-known constant from the equation for the critical power of relativistic self-focussing of a laser pulse in plasma \cite{Litvak_1970_JETP_30_166, Sprangle_1987_IEEETPS_15_145},
\begin{equation}
    P_\critical = \frac{2 m^2 c^5}{e^2} \frac{\omega_\laser^2}{\omega_\plasm} \approx \frac{\omega_\laser^2}{\omega_\plasm^2}\times \SI{17.4}{GW}.
\end{equation}

First of all, using Eq.~\eqref{eq:I0Physical} we can directly link the value of $I_0$ to the size of the bubble,
\begin{equation}
    I_0[\si{GW}] \approx 212 \qty(\frac{\Rb}{\lambda_\plasm})^4,
\end{equation}
where $\lambda_\plasm = 2\pi c / \omega_\plasm$ is the plasma wavelength.
Typically, the size of the bubble is not much different from the wavelength, but the value of $I_0$ is very sensitive to small changes in the size.

\subsubsection{Power of the bubble generated by an electron driver}

Next, we rewrite Eq.~\eqref{eq:analytical:I0_analytical_Q} in physical units.
Note that the total charge $Q$ in our units is normalized not to $e$, but to $e n_\plasm k_\plasm^{-3}$.
In this case, in Gaussian units,
\begin{equation}
    I_0 = \frac{2 \pi m c^3}{\lambda_\plasm} \frac{Q}{e} \left[\sqrt{\frac{2 e Q}{m c^2 \xi_\bunch}} - \frac{\pi}{4} \frac{\xi_\bunch}{\lambda_\plasm}\right].
    \label{eq:scaling:I0_dimensional}
\end{equation}
If we evaluate the constants in this formula, then
\begin{equation}
    I_0[\si{GW}] \approx 963 \frac{Q[\si{nC}]}{\lambda_\plasm[\si{\um}] } \left[\sqrt{35.18 \frac{Q[\si{nC}]}{\xi_\bunch[\si{\um}]}} - \frac{\pi}{4} \frac{\xi_\bunch}{\lambda_\plasm} \right].
\end{equation}
At the specified total charge, the maximum possible length of the bunch (at which $I_0$ reaches 0) for the specified charge is
\begin{align}
    &\xi_{\bunch,\submax} = \qty(\frac{32 e Q \lambda_\plasm^2}{\pi^2 m c^2})^{1/3},\\
    &\xi_{\bunch,\submax}[\si{\um}] \approx 3.85 Q^{1/3}[\si{nC}] \lambda_\plasm^{2/3}[\si{\um}].
\end{align}
The maximum length of the driver fitting in a bubble drops with the decrease of the plasma wavelength (increase of the plasma density).
If $\xi_\bunch \ll \xi_{\bunch, \submax}$, the estimate for $I_0$ becomes
\begin{align}
    &I_0 = 2\pi c^2 \sqrt{\frac{2m}{e}} \frac{Q^{3/2}}{\xi_\bunch^{1/2} \lambda_\plasm},\\
    &I_0[\si{GW}] \approx 5709 \frac{Q^{3/2}[\si{nC}]}{\xi_\bunch^{1/2}[\si{\um}] \lambda_\plasm[\si{\um}]}.
\end{align}
This shows that the shorter the bunch, the more powerful bubble it generates.
The model does not limit the maximum possible value of $I_0$, but the model applicability will likely fail for extremely short bunches.

\subsubsection{Optimal plasma density for the driver}

If we have a driver with fixed parameters (charge and length), then the value of $I_0$ depends only on the plasma density.
From Eq.~\eqref{eq:scaling:I0_dimensional} we can conclude that for very large plasma densities (small $\lambda_\plasm$) we technically have $I_0 < 0$, which means that the bunch is too large to fully fit in the bubble it excites.
For very small plasma densities, on the contrary, $I_0$ becomes inversely proportional to $\lambda_\plasm$ or proportional to $n_\plasm^{1/2}$.
Thus, the optimal value of plasma density exists which leads to the maximum value of $I_0$.
This value corresponds to
\begin{equation}
    \lambda_{\plasm,\opt} = \frac{\pi}{2} \sqrt{\frac{m c^2 \xi_\bunch^3}{2 e Q}}, \quad n_{\plasm,\opt} = \frac{8}{\pi \xi_\bunch^3} \frac{Q}{e},
    \label{eq:scaling:n_opt}
\end{equation}
or, in physical units,
\begin{equation}
    n_{\plasm,\opt}[\si{cm^{-3}}] = \num{1.59e21} \frac{Q[\si{nC}]}{\xi_\bunch^3[\si{\um}]}.
\end{equation}
The corresponding maximum value of $I_0$ is 
\begin{equation}
    I_{0,\submax} = \frac{4 Q^2 c}{\xi_\bunch^2}, \quad I_{0,\submax}[\si{GW}] \approx 10778 \frac{Q^2[\si{nC}]}{\xi_\bunch^2[\si{\um}]}.
\end{equation}

As the value of $I_0$ determines the maximum power of acceleration, it might be beneficial to operate in this optimal regime.
The scaling again shows that shortening the driver as much as possible is required.
However, it is also important that the dependence on the charge $Q$ is quadratic, which means that increasing the total charge of the driver can vastly increase the potential of the plasma accelerator.

\subsubsection{Deceleration length of the driver}

The power of the bubble $I_0$ is also equal to the rate at which the driver loses its energy.
This allows us to quantitatively estimate the deceleration length of the bunch, i.e. the distance after which the driver completely depletes its energy.
The kinetic energy of the bunch with the average Lorentz factor of $\gamma$ is
\begin{equation}
    K_\bunch = \frac{Q}{e} K_\electron = \frac{Q}{e} (\gamma - 1) mc^2,
\end{equation}
where $K_\electron$ is the average kinetic energy of a single electron.
It is convenient to measure it the units of MeV.
In this case, the deceleration length is
\begin{equation}
    L_\deceleration = c\frac{K_\bunch}{I_0} = \frac{K_\electron}{mc^2} \frac{\lambda_\plasm}{2\pi} \left[\sqrt{\frac{2 e Q}{m c^2 \xi_\bunch}} - \frac{\pi}{4} \frac{\xi_\bunch}{\lambda_\plasm}\right]^{-1}.
\end{equation}
For the optimal plasma density \eqref{eq:scaling:n_opt}, the deceleration length is the shortest possible for the specified driver,
\begin{align}
    &L_{\deceleration,\submin} = \frac{K_\electron \xi_\bunch^2}{4 e Q}, \\
    &L_{\deceleration,\submin}[\si{cm}] \approx \num{2.78e-6} \frac{K_\electron[\si{MeV}] \xi_\bunch^2[\si{\um}]}{Q[\si{nC}]}.
\end{align}

Obviously, as the bunch is located in the non-uniform decelerating field as well as the transverse field, this serves as a rough estimate, because the bunch will change its shape and break before it fully depletes.

\section*{Discussion}

In this paper, we presented an analytical model which makes it possible to calculate the excitation of the bubble based on the properties of the driver.
The model is based on solving Eq.~\eqref{eq:bubbleEquationUniform} in the relativistic approximation with $\rb = 0$ as the initial condition.
As the second derivative in this equation diverges at $\rb = 0$, in order to be able to calculate the solution numerically, we propose making a substitution \eqref{eq:psiEquation} and solving Eq.~\eqref{eq:psiEquation} with initial conditions \eqref{eq:psiEquationInitialConditions} instead.

Similar calculations could be done using full Eq.~\eqref{eq:bubbleEquationGeneral} without any approximations.
In this case, it does not possess any singularities at $\rb = 0$ (e.g., see Ref.~\cite{Lu_2006_PRL_96_165002}) and can be easily integrated with $\rb=0$ as the initial condition.
However, the coefficients in this equation depend on some external parameters such as the shape of the electron sheath on the boundary of the bubble and its width $\Delta$.
In this sense, this solution is also not self-consistent, as we have to set these parameters manually in order to achieve good correspondence to the simulation results.
Moreover, Eq.~\eqref{eq:bubbleEquationGeneral} usually does not describe the electric field in the rear part of the bubble well and requires corrections \cite{Reichwein_2020_PPCF_62_115017}.
Our approach does not have such a drawback: the solution is determined solely by the properties of the driver, and it provides fairly good estimates for the electric field in the bubble, as shown by Figs.~\ref{fig:bubbleExample}, \ref{fig:bubbles-gaussian}.
It is interesting that this approach leads to a very good correspondence even in the region $\rb \lesssim 1$ where the approximations used in our models should not be valid.
This perhaps indicates that Eq.~\eqref{eq:bubbleEquationUniform} in the relativistic approximation has a broader area of applicability than follows from our estimates, although the complete understanding on why this might be the case is currently lacking.

We also presented a case of a cylindrical driver when the equation can be solved analytically.
Using this case, we were able to find analytical scaling laws for PWFA in the blowout regime.

Another important question for future research is whether similar approach can be applied for laser driven strongly nonlinear wakes, as LWFA is significantly more interesting for practical purposes.
However, even though a laser-driven bubble can be described with Eq.~\eqref{eq:bubbleEquationUniform} too in the areas where the laser field is absent, the bubble does not form such a well-defined sheath in the area inside the laser pulse \cite{Lu_2006_PRL_96_165002}, which can limit the applicability of the used model.

Our model is also not currently applicable to positively charged particle drivers (e.g., proton or positron), because the physics of the excitation for them is qualitatively different.
The driver first attracts the electrons of the plasma towards the axis of its propagation, and then these electrons overshoot and create a blowout.
Creating such a model is also a topic for future research.

\begin{acknowledgements}

The research has been supported by RFBR and DFG (project No.~20-52-12046), by RFBR (Project No.~20-02-00691), and by BMBF (Project 05K16PFB).

\end{acknowledgements}

\appendix
\section{Initial stage of excitation} \label{appendix:analyticalExcitation}

As we have shown in Section~\ref{sec:excitation}, the solution to Eq.~\eqref{eq:psiEquation} should be started from the point where the charge density of the driver $\abs{\rho_\bunch}$ reaches the value of $1/2$.
Around this point, for sufficiently small deviations, every charge density distribution can be approximated by a linear function in $\xi$, and its dependence on the transverse coordinate can be neglected,
\begin{equation}
    \rho_\bunch(\xi, r) \approx -\frac{1}{2} - \rho_0 \frac{\xi}{L}.
\end{equation}
In this case, according to Eq.~\eqref{eq:kappa},
\begin{equation}
    \varkappa(\xi, r) = \frac{1}{2} + \rho_0 \frac{\xi}{L},
\end{equation}
and the solution to Eq.~\eqref{eq:psiEquation} is
\begin{equation}
    \psi_\xi = \frac{2}{21} \frac{\rho_0}{L} \xi^3.
\end{equation}
The condition that the second term of Eq.~\eqref{eq:psiEquation} turns to 0 is satisfied.

If we write other quantities, we get
\begin{equation}
    \rb(\xi) = \sqrt{\frac{8}{21}\frac{\rho_0}{L}} \xi^{3/2}, \quad E_z(\xi) = \frac{2}{7} \frac{\rho_0}{L} \xi^2.
\end{equation}
Here, $\dv*[2]{\rb}{\xi} \propto \xi^{-1/2}$ diverges at $0$, which shows we cannot use Eq.~\eqref{eq:bubbleEquationUniform} to find this solution directly.

\section{PIC simulation details} \label{appendix:simulations}

PIC simulations in the paper were performed using the Smilei PIC code \cite{Derouillat_2018_CPC_222_351, Smilei}.

For simulations in Fig.~\ref{fig:bubbleExample}, both the driver and the witness had the parabolic profile
\begin{equation}
    \rho(\xi, r) = -\rho_0 \left[1 - \frac{(\xi - \xi_0)^2}{\xi_\sigma^2} \right] \left[1 - \frac{r^2}{r_\sigma^2} \right]
\end{equation}
with $\xi_\sigma = 2$, $r_\sigma = 1$, and different values of $\rho_0$ equal to $100$ (driver) and $50$ (witness).
The simulation box was $30 \times 30 \times 25$ (in plasma units $k_\plasm^{-1}$) with spatial steps of $0.1$ for all three dimensions.
The number of particles per cell was equal to $8$.

For simulations in Fig.~\ref{fig:bubbles-gaussian}, the driver had the Gaussian shape
\begin{equation}
    \rho(\xi, r) = - \rho_0 \exp[- \frac{(\xi - \xi_0)^2}{\xi_\sigma^2} - \frac{r^2}{r_\sigma^2}]
\end{equation}
with $\xi_\sigma = 2$, $r_\sigma = 1$.
The simulation box was $30 \times 30 \times 35$ with spatial steps of $0.2$ for all three dimensions.
The number of particles per cell was equal to 1.

In all simulations, the driver and witness bunches had a velocity corresponding to the energy of \SI{2}{GeV}, but the mass of the electrons in the bunches was artificially increased to a very large value to prevent their betatron oscillations and make the comparison with the theoretical results more straightforward.
The moving window technique was used to follow the driver.
The snapshots of the wakefields were taken at the moment of time where the wakefields were fully established and stopped evolving after the bunch transition through the abrupt vacuum--plasma border.

\bibliographystyle{aipnum4-1}
\bibliography{Bibliography}

\begin{thebibliography}{35}%
\makeatletter
\providecommand \@ifxundefined [1]{%
 \@ifx{#1\undefined}
}%
\providecommand \@ifnum [1]{%
 \ifnum #1\expandafter \@firstoftwo
 \else \expandafter \@secondoftwo
 \fi
}%
\providecommand \@ifx [1]{%
 \ifx #1\expandafter \@firstoftwo
 \else \expandafter \@secondoftwo
 \fi
}%
\providecommand \natexlab [1]{#1}%
\providecommand \enquote  [1]{``#1''}%
\providecommand \bibnamefont  [1]{#1}%
\providecommand \bibfnamefont [1]{#1}%
\providecommand \citenamefont [1]{#1}%
\providecommand \href@noop [0]{\@secondoftwo}%
\providecommand \href [0]{\begingroup \@sanitize@url \@href}%
\providecommand \@href[1]{\@@startlink{#1}\@@href}%
\providecommand \@@href[1]{\endgroup#1\@@endlink}%
\providecommand \@sanitize@url [0]{\catcode `\\12\catcode `\$12\catcode
  `\&12\catcode `\#12\catcode `\^12\catcode `\_12\catcode `\%12\relax}%
\providecommand \@@startlink[1]{}%
\providecommand \@@endlink[0]{}%
\providecommand \url  [0]{\begingroup\@sanitize@url \@url }%
\providecommand \@url [1]{\endgroup\@href {#1}{\urlprefix }}%
\providecommand \urlprefix  [0]{URL }%
\providecommand \Eprint [0]{\href }%
\providecommand \doibase [0]{http://dx.doi.org/}%
\providecommand \selectlanguage [0]{\@gobble}%
\providecommand \bibinfo  [0]{\@secondoftwo}%
\providecommand \bibfield  [0]{\@secondoftwo}%
\providecommand \translation [1]{[#1]}%
\providecommand \BibitemOpen [0]{}%
\providecommand \bibitemStop [0]{}%
\providecommand \bibitemNoStop [0]{.\EOS\space}%
\providecommand \EOS [0]{\spacefactor3000\relax}%
\providecommand \BibitemShut  [1]{\csname bibitem#1\endcsname}%
\let\auto@bib@innerbib\@empty
\bibitem [{\citenamefont {Esarey}, \citenamefont {Schroeder},\ and\
  \citenamefont {Leemans}(2009)}]{Esarey_2009_RMP_81_1229}%
  \BibitemOpen
  \bibfield  {author} {\bibinfo {author} {\bibfnamefont {E.}~\bibnamefont
  {Esarey}}, \bibinfo {author} {\bibfnamefont {C.~B.}\ \bibnamefont
  {Schroeder}}, \ and\ \bibinfo {author} {\bibfnamefont {W.~P.}\ \bibnamefont
  {Leemans}},\ }\href {\doibase 10.1103/RevModPhys.81.1229} {\bibfield
  {journal} {\bibinfo  {journal} {Rev. Mod. Phys.}\ }\textbf {\bibinfo {volume}
  {81}},\ \bibinfo {pages} {1229} (\bibinfo {year} {2009})}\BibitemShut
  {NoStop}%
\bibitem [{\citenamefont {Kostyukov}\ and\ \citenamefont
  {Pukhov}(2015)}]{Kostyukov_2015_UFN_58_81}%
  \BibitemOpen
  \bibfield  {author} {\bibinfo {author} {\bibfnamefont {I.~{\relax Yu}.}\
  \bibnamefont {Kostyukov}}\ and\ \bibinfo {author} {\bibfnamefont {A.~M.}\
  \bibnamefont {Pukhov}},\ }\href {\doibase 10.3367/ufne.0185.201501g.0089}
  {\bibfield  {journal} {\bibinfo  {journal} {Phys.-Uspekhi}\ }\textbf
  {\bibinfo {volume} {58}},\ \bibinfo {pages} {81} (\bibinfo {year}
  {2015})}\BibitemShut {NoStop}%
\bibitem [{\citenamefont {Tajima}\ and\ \citenamefont
  {Dawson}(1979)}]{Tajima_1979_PRL_43_267}%
  \BibitemOpen
  \bibfield  {author} {\bibinfo {author} {\bibfnamefont {T.}~\bibnamefont
  {Tajima}}\ and\ \bibinfo {author} {\bibfnamefont {J.~M.}\ \bibnamefont
  {Dawson}},\ }\href {\doibase 10.1103/PhysRevLett.43.267} {\bibfield
  {journal} {\bibinfo  {journal} {Phys. Rev. Lett.}\ }\textbf {\bibinfo
  {volume} {43}},\ \bibinfo {pages} {267} (\bibinfo {year} {1979})}\BibitemShut
  {NoStop}%
\bibitem [{\citenamefont {Chen}\ \emph {et~al.}(1985)\citenamefont {Chen},
  \citenamefont {Dawson}, \citenamefont {Huff},\ and\ \citenamefont
  {Katsouleas}}]{Chen_1985_PRL_54_693}%
  \BibitemOpen
  \bibfield  {author} {\bibinfo {author} {\bibfnamefont {P.}~\bibnamefont
  {Chen}}, \bibinfo {author} {\bibfnamefont {J.~M.}\ \bibnamefont {Dawson}},
  \bibinfo {author} {\bibfnamefont {R.~W.}\ \bibnamefont {Huff}}, \ and\
  \bibinfo {author} {\bibfnamefont {T.}~\bibnamefont {Katsouleas}},\ }\href
  {\doibase 10.1103/PhysRevLett.54.693} {\bibfield  {journal} {\bibinfo
  {journal} {Phys. Rev. Lett.}\ }\textbf {\bibinfo {volume} {54}},\ \bibinfo
  {pages} {693} (\bibinfo {year} {1985})}\BibitemShut {NoStop}%
\bibitem [{\citenamefont {Rosenzweig}\ \emph {et~al.}(1988)\citenamefont
  {Rosenzweig}, \citenamefont {Cline}, \citenamefont {Cole}, \citenamefont
  {Figueroa}, \citenamefont {Gai}, \citenamefont {Konecny}, \citenamefont
  {Norem}, \citenamefont {Schoessow},\ and\ \citenamefont
  {Simpson}}]{Rosenzweig_1988_PRL_61_98}%
  \BibitemOpen
  \bibfield  {author} {\bibinfo {author} {\bibfnamefont {J.~B.}\ \bibnamefont
  {Rosenzweig}}, \bibinfo {author} {\bibfnamefont {D.~B.}\ \bibnamefont
  {Cline}}, \bibinfo {author} {\bibfnamefont {B.}~\bibnamefont {Cole}},
  \bibinfo {author} {\bibfnamefont {H.}~\bibnamefont {Figueroa}}, \bibinfo
  {author} {\bibfnamefont {W.}~\bibnamefont {Gai}}, \bibinfo {author}
  {\bibfnamefont {R.}~\bibnamefont {Konecny}}, \bibinfo {author} {\bibfnamefont
  {J.}~\bibnamefont {Norem}}, \bibinfo {author} {\bibfnamefont
  {P.}~\bibnamefont {Schoessow}}, \ and\ \bibinfo {author} {\bibfnamefont
  {J.}~\bibnamefont {Simpson}},\ }\href {\doibase 10.1103/PhysRevLett.61.98}
  {\bibfield  {journal} {\bibinfo  {journal} {Phys. Rev. Lett.}\ }\textbf
  {\bibinfo {volume} {61}},\ \bibinfo {pages} {98} (\bibinfo {year}
  {1988})}\BibitemShut {NoStop}%
\bibitem [{\citenamefont {Quesnel}\ and\ \citenamefont
  {Mora}(1998)}]{Quesnel_1998_PRE_58_3719}%
  \BibitemOpen
  \bibfield  {author} {\bibinfo {author} {\bibfnamefont {B.}~\bibnamefont
  {Quesnel}}\ and\ \bibinfo {author} {\bibfnamefont {P.}~\bibnamefont {Mora}},\
  }\href {\doibase 10.1103/PhysRevE.58.3719} {\bibfield  {journal} {\bibinfo
  {journal} {Phys. Rev. E}\ }\textbf {\bibinfo {volume} {58}},\ \bibinfo
  {pages} {3719} (\bibinfo {year} {1998})}\BibitemShut {NoStop}%
\bibitem [{\citenamefont {Fainberg}(1968)}]{Fainberg_1968_PhysUsp_10_750}%
  \BibitemOpen
  \bibfield  {author} {\bibinfo {author} {\bibfnamefont {Y.~B.}\ \bibnamefont
  {Fainberg}},\ }\href {\doibase 10.1070/PU1968v010n06ABEH003715} {\bibfield
  {journal} {\bibinfo  {journal} {Sov. Phys. Usp.}\ }\textbf {\bibinfo {volume}
  {10}},\ \bibinfo {pages} {750} (\bibinfo {year} {1968})}\BibitemShut
  {NoStop}%
\bibitem [{\citenamefont {Strickland}\ and\ \citenamefont
  {Mourou}(1985)}]{Strickland1985compression}%
  \BibitemOpen
  \bibfield  {author} {\bibinfo {author} {\bibfnamefont {D.}~\bibnamefont
  {Strickland}}\ and\ \bibinfo {author} {\bibfnamefont {G.}~\bibnamefont
  {Mourou}},\ }\href {\doibase 10.1016/0030-4018(85)90151-8} {\bibfield
  {journal} {\bibinfo  {journal} {Opt. Commun.}\ }\textbf {\bibinfo {volume}
  {55}},\ \bibinfo {pages} {447} (\bibinfo {year} {1985})}\BibitemShut
  {NoStop}%
\bibitem [{\citenamefont {Pukhov}\ and\ \citenamefont
  {{Meyer-ter-Vehn}}(2002)}]{Pukhov2002Bubble}%
  \BibitemOpen
  \bibfield  {author} {\bibinfo {author} {\bibfnamefont {A.}~\bibnamefont
  {Pukhov}}\ and\ \bibinfo {author} {\bibfnamefont {J.}~\bibnamefont
  {{Meyer-ter-Vehn}}},\ }\href {\doibase 10.1007/s003400200795} {\bibfield
  {journal} {\bibinfo  {journal} {Appl. Phys. B}\ }\textbf {\bibinfo {volume}
  {74}},\ \bibinfo {pages} {355} (\bibinfo {year} {2002})}\BibitemShut
  {NoStop}%
\bibitem [{\citenamefont {Gonsalves}\ \emph {et~al.}(2019)\citenamefont
  {Gonsalves}, \citenamefont {Nakamura}, \citenamefont {Daniels}, \citenamefont
  {Benedetti}, \citenamefont {Pieronek}, \citenamefont {de~Raadt},
  \citenamefont {Steinke}, \citenamefont {Bin}, \citenamefont {Bulanov},
  \citenamefont {van Tilborg}, \citenamefont {Geddes}, \citenamefont
  {Schroeder}, \citenamefont {T\'oth}, \citenamefont {Esarey}, \citenamefont
  {Swanson}, \citenamefont {Fan-Chiang}, \citenamefont {Bagdasarov},
  \citenamefont {Bobrova}, \citenamefont {Gasilov}, \citenamefont {Korn},
  \citenamefont {Sasorov},\ and\ \citenamefont
  {Leemans}}]{Gonsalves_2019_PRL_122_084801}%
  \BibitemOpen
  \bibfield  {author} {\bibinfo {author} {\bibfnamefont {A.~J.}\ \bibnamefont
  {Gonsalves}}, \bibinfo {author} {\bibfnamefont {K.}~\bibnamefont {Nakamura}},
  \bibinfo {author} {\bibfnamefont {J.}~\bibnamefont {Daniels}}, \bibinfo
  {author} {\bibfnamefont {C.}~\bibnamefont {Benedetti}}, \bibinfo {author}
  {\bibfnamefont {C.}~\bibnamefont {Pieronek}}, \bibinfo {author}
  {\bibfnamefont {T.~C.~H.}\ \bibnamefont {de~Raadt}}, \bibinfo {author}
  {\bibfnamefont {S.}~\bibnamefont {Steinke}}, \bibinfo {author} {\bibfnamefont
  {J.~H.}\ \bibnamefont {Bin}}, \bibinfo {author} {\bibfnamefont {S.~S.}\
  \bibnamefont {Bulanov}}, \bibinfo {author} {\bibfnamefont {J.}~\bibnamefont
  {van Tilborg}}, \bibinfo {author} {\bibfnamefont {C.~G.~R.}\ \bibnamefont
  {Geddes}}, \bibinfo {author} {\bibfnamefont {C.~B.}\ \bibnamefont
  {Schroeder}}, \bibinfo {author} {\bibfnamefont {C.}~\bibnamefont {T\'oth}},
  \bibinfo {author} {\bibfnamefont {E.}~\bibnamefont {Esarey}}, \bibinfo
  {author} {\bibfnamefont {K.}~\bibnamefont {Swanson}}, \bibinfo {author}
  {\bibfnamefont {L.}~\bibnamefont {Fan-Chiang}}, \bibinfo {author}
  {\bibfnamefont {G.}~\bibnamefont {Bagdasarov}}, \bibinfo {author}
  {\bibfnamefont {N.}~\bibnamefont {Bobrova}}, \bibinfo {author} {\bibfnamefont
  {V.}~\bibnamefont {Gasilov}}, \bibinfo {author} {\bibfnamefont
  {G.}~\bibnamefont {Korn}}, \bibinfo {author} {\bibfnamefont {P.}~\bibnamefont
  {Sasorov}}, \ and\ \bibinfo {author} {\bibfnamefont {W.~P.}\ \bibnamefont
  {Leemans}},\ }\href {\doibase 10.1103/PhysRevLett.122.084801} {\bibfield
  {journal} {\bibinfo  {journal} {Phys. Rev. Lett.}\ }\textbf {\bibinfo
  {volume} {122}},\ \bibinfo {pages} {084801} (\bibinfo {year}
  {2019})}\BibitemShut {NoStop}%
\bibitem [{\citenamefont {Rosenzweig}\ \emph {et~al.}(1991)\citenamefont
  {Rosenzweig}, \citenamefont {Breizman}, \citenamefont {Katsouleas},\ and\
  \citenamefont {Su}}]{Rosenzweig_1991_PRA_44_R6189}%
  \BibitemOpen
  \bibfield  {author} {\bibinfo {author} {\bibfnamefont {J.~B.}\ \bibnamefont
  {Rosenzweig}}, \bibinfo {author} {\bibfnamefont {B.}~\bibnamefont
  {Breizman}}, \bibinfo {author} {\bibfnamefont {T.}~\bibnamefont
  {Katsouleas}}, \ and\ \bibinfo {author} {\bibfnamefont {J.~J.}\ \bibnamefont
  {Su}},\ }\href {\doibase 10.1103/PhysRevA.44.R6189} {\bibfield  {journal}
  {\bibinfo  {journal} {Phys. Rev. A}\ }\textbf {\bibinfo {volume} {44}},\
  \bibinfo {pages} {R6189} (\bibinfo {year} {1991})}\BibitemShut {NoStop}%
\bibitem [{\citenamefont {Aschikhin}\ \emph {et~al.}(2016)\citenamefont
  {Aschikhin}, \citenamefont {Behrens}, \citenamefont {Bohlen}, \citenamefont
  {Dale}, \citenamefont {Delbos}, \citenamefont {{di Lucchio}}, \citenamefont
  {Elsen}, \citenamefont {Erbe}, \citenamefont {Felber}, \citenamefont
  {Foster}, \citenamefont {Goldberg}, \citenamefont {Grebenyuk}, \citenamefont
  {Gruse}, \citenamefont {Hidding}, \citenamefont {Hu}, \citenamefont
  {Karstensen}, \citenamefont {Knetsch}, \citenamefont {Kononenko},
  \citenamefont {Libov}, \citenamefont {Ludwig}, \citenamefont {Maier},
  \citenamefont {{Martinez de la Ossa}}, \citenamefont {Mehrling},
  \citenamefont {Palmer}, \citenamefont {Pannek}, \citenamefont {Schaper},
  \citenamefont {Schlarb}, \citenamefont {Schmidt}, \citenamefont {Schreiber},
  \citenamefont {Schwinkendorf}, \citenamefont {Steel}, \citenamefont
  {Streeter}, \citenamefont {Tauscher}, \citenamefont {Wacker}, \citenamefont
  {Weichert}, \citenamefont {Wunderlich}, \citenamefont {Zemella},\ and\
  \citenamefont {Osterhoff}}]{Aschikhin_2016_NIMA_806_175}%
  \BibitemOpen
  \bibfield  {author} {\bibinfo {author} {\bibfnamefont {A.}~\bibnamefont
  {Aschikhin}}, \bibinfo {author} {\bibfnamefont {C.}~\bibnamefont {Behrens}},
  \bibinfo {author} {\bibfnamefont {S.}~\bibnamefont {Bohlen}}, \bibinfo
  {author} {\bibfnamefont {J.}~\bibnamefont {Dale}}, \bibinfo {author}
  {\bibfnamefont {N.}~\bibnamefont {Delbos}}, \bibinfo {author} {\bibfnamefont
  {L.}~\bibnamefont {{di Lucchio}}}, \bibinfo {author} {\bibfnamefont
  {E.}~\bibnamefont {Elsen}}, \bibinfo {author} {\bibfnamefont {J.-H.}\
  \bibnamefont {Erbe}}, \bibinfo {author} {\bibfnamefont {M.}~\bibnamefont
  {Felber}}, \bibinfo {author} {\bibfnamefont {B.}~\bibnamefont {Foster}},
  \bibinfo {author} {\bibfnamefont {L.}~\bibnamefont {Goldberg}}, \bibinfo
  {author} {\bibfnamefont {J.}~\bibnamefont {Grebenyuk}}, \bibinfo {author}
  {\bibfnamefont {J.-N.}\ \bibnamefont {Gruse}}, \bibinfo {author}
  {\bibfnamefont {B.}~\bibnamefont {Hidding}}, \bibinfo {author} {\bibfnamefont
  {Z.}~\bibnamefont {Hu}}, \bibinfo {author} {\bibfnamefont {S.}~\bibnamefont
  {Karstensen}}, \bibinfo {author} {\bibfnamefont {A.}~\bibnamefont {Knetsch}},
  \bibinfo {author} {\bibfnamefont {O.}~\bibnamefont {Kononenko}}, \bibinfo
  {author} {\bibfnamefont {V.}~\bibnamefont {Libov}}, \bibinfo {author}
  {\bibfnamefont {K.}~\bibnamefont {Ludwig}}, \bibinfo {author} {\bibfnamefont
  {A.}~\bibnamefont {Maier}}, \bibinfo {author} {\bibfnamefont
  {A.}~\bibnamefont {{Martinez de la Ossa}}}, \bibinfo {author} {\bibfnamefont
  {T.}~\bibnamefont {Mehrling}}, \bibinfo {author} {\bibfnamefont
  {C.}~\bibnamefont {Palmer}}, \bibinfo {author} {\bibfnamefont
  {F.}~\bibnamefont {Pannek}}, \bibinfo {author} {\bibfnamefont
  {L.}~\bibnamefont {Schaper}}, \bibinfo {author} {\bibfnamefont
  {H.}~\bibnamefont {Schlarb}}, \bibinfo {author} {\bibfnamefont
  {B.}~\bibnamefont {Schmidt}}, \bibinfo {author} {\bibfnamefont
  {S.}~\bibnamefont {Schreiber}}, \bibinfo {author} {\bibfnamefont {J.-P.}\
  \bibnamefont {Schwinkendorf}}, \bibinfo {author} {\bibfnamefont
  {H.}~\bibnamefont {Steel}}, \bibinfo {author} {\bibfnamefont
  {M.}~\bibnamefont {Streeter}}, \bibinfo {author} {\bibfnamefont
  {G.}~\bibnamefont {Tauscher}}, \bibinfo {author} {\bibfnamefont
  {V.}~\bibnamefont {Wacker}}, \bibinfo {author} {\bibfnamefont
  {S.}~\bibnamefont {Weichert}}, \bibinfo {author} {\bibfnamefont
  {S.}~\bibnamefont {Wunderlich}}, \bibinfo {author} {\bibfnamefont
  {J.}~\bibnamefont {Zemella}}, \ and\ \bibinfo {author} {\bibfnamefont
  {J.}~\bibnamefont {Osterhoff}},\ }\href {\doibase
  https://doi.org/10.1016/j.nima.2015.10.005} {\bibfield  {journal} {\bibinfo
  {journal} {Nucl. Instrum. Meth. A}\ }\textbf {\bibinfo {volume} {806}},\
  \bibinfo {pages} {175} (\bibinfo {year} {2016})}\BibitemShut {NoStop}%
\bibitem [{\citenamefont {Martinez de~la Ossa}\ \emph
  {et~al.}(2019)\citenamefont {Martinez de~la Ossa}, \citenamefont {Assmann},
  \citenamefont {Bussmann}, \citenamefont {Corde}, \citenamefont
  {Couperus~Cabadağ}, \citenamefont {Debus}, \citenamefont {Döpp},
  \citenamefont {Ferran~Pousa}, \citenamefont {Gilljohann}, \citenamefont
  {Heinemann}, \citenamefont {Hidding}, \citenamefont {Irman}, \citenamefont
  {Karsch}, \citenamefont {Kononenko}, \citenamefont {Kurz}, \citenamefont
  {Osterhoff}, \citenamefont {Pausch}, \citenamefont {Schöbel},\ and\
  \citenamefont {Schramm}}]{Martinez_2019_PTRA_377_20180175}%
  \BibitemOpen
  \bibfield  {author} {\bibinfo {author} {\bibfnamefont {A.}~\bibnamefont
  {Martinez de~la Ossa}}, \bibinfo {author} {\bibfnamefont {R.~W.}\
  \bibnamefont {Assmann}}, \bibinfo {author} {\bibfnamefont {M.}~\bibnamefont
  {Bussmann}}, \bibinfo {author} {\bibfnamefont {S.}~\bibnamefont {Corde}},
  \bibinfo {author} {\bibfnamefont {J.~P.}\ \bibnamefont {Couperus~Cabadağ}},
  \bibinfo {author} {\bibfnamefont {A.}~\bibnamefont {Debus}}, \bibinfo
  {author} {\bibfnamefont {A.}~\bibnamefont {Döpp}}, \bibinfo {author}
  {\bibfnamefont {A.}~\bibnamefont {Ferran~Pousa}}, \bibinfo {author}
  {\bibfnamefont {M.~F.}\ \bibnamefont {Gilljohann}}, \bibinfo {author}
  {\bibfnamefont {T.}~\bibnamefont {Heinemann}}, \bibinfo {author}
  {\bibfnamefont {B.}~\bibnamefont {Hidding}}, \bibinfo {author} {\bibfnamefont
  {A.}~\bibnamefont {Irman}}, \bibinfo {author} {\bibfnamefont
  {S.}~\bibnamefont {Karsch}}, \bibinfo {author} {\bibfnamefont
  {O.}~\bibnamefont {Kononenko}}, \bibinfo {author} {\bibfnamefont
  {T.}~\bibnamefont {Kurz}}, \bibinfo {author} {\bibfnamefont {J.}~\bibnamefont
  {Osterhoff}}, \bibinfo {author} {\bibfnamefont {R.}~\bibnamefont {Pausch}},
  \bibinfo {author} {\bibfnamefont {S.}~\bibnamefont {Schöbel}}, \ and\
  \bibinfo {author} {\bibfnamefont {U.}~\bibnamefont {Schramm}},\ }\href
  {\doibase 10.1098/rsta.2018.0175} {\bibfield  {journal} {\bibinfo  {journal}
  {Philos. Trans. R. Soc A}\ }\textbf {\bibinfo {volume} {377}},\ \bibinfo
  {pages} {20180175} (\bibinfo {year} {2019})}\BibitemShut {NoStop}%
\bibitem [{\citenamefont {Ferri}\ \emph {et~al.}(2018)\citenamefont {Ferri},
  \citenamefont {Corde}, \citenamefont {D\"opp}, \citenamefont {Lifschitz},
  \citenamefont {Doche}, \citenamefont {Thaury}, \citenamefont {Ta~Phuoc},
  \citenamefont {Mahieu}, \citenamefont {Andriyash}, \citenamefont {Malka},\
  and\ \citenamefont {Davoine}}]{Ferri_2018_PRL_120_254802}%
  \BibitemOpen
  \bibfield  {author} {\bibinfo {author} {\bibfnamefont {J.}~\bibnamefont
  {Ferri}}, \bibinfo {author} {\bibfnamefont {S.}~\bibnamefont {Corde}},
  \bibinfo {author} {\bibfnamefont {A.}~\bibnamefont {D\"opp}}, \bibinfo
  {author} {\bibfnamefont {A.}~\bibnamefont {Lifschitz}}, \bibinfo {author}
  {\bibfnamefont {A.}~\bibnamefont {Doche}}, \bibinfo {author} {\bibfnamefont
  {C.}~\bibnamefont {Thaury}}, \bibinfo {author} {\bibfnamefont
  {K.}~\bibnamefont {Ta~Phuoc}}, \bibinfo {author} {\bibfnamefont
  {B.}~\bibnamefont {Mahieu}}, \bibinfo {author} {\bibfnamefont {I.~A.}\
  \bibnamefont {Andriyash}}, \bibinfo {author} {\bibfnamefont {V.}~\bibnamefont
  {Malka}}, \ and\ \bibinfo {author} {\bibfnamefont {X.}~\bibnamefont
  {Davoine}},\ }\href {\doibase 10.1103/PhysRevLett.120.254802} {\bibfield
  {journal} {\bibinfo  {journal} {Phys. Rev. Lett.}\ }\textbf {\bibinfo
  {volume} {120}},\ \bibinfo {pages} {254802} (\bibinfo {year}
  {2018})}\BibitemShut {NoStop}%
\bibitem [{\citenamefont {Gschwendtner}\ \emph {et~al.}(2019)\citenamefont
  {Gschwendtner}, \citenamefont {Turner}, \citenamefont {Adli}, \citenamefont
  {Ahuja}, \citenamefont {Apsimon}, \citenamefont {Apsimon}, \citenamefont
  {Bachmann}, \citenamefont {Batsch}, \citenamefont {Bracco}, \citenamefont
  {Braunmüller}, \citenamefont {Burger}, \citenamefont {Burt}, \citenamefont
  {Buttenschön}, \citenamefont {Caldwell}, \citenamefont {Chappell},
  \citenamefont {Chevallay}, \citenamefont {Chung}, \citenamefont {Cooke},
  \citenamefont {Damerau}, \citenamefont {Deubner}, \citenamefont {Dexter},
  \citenamefont {Doebert}, \citenamefont {Farmer}, \citenamefont {Fedosseev},
  \citenamefont {Fiorito}, \citenamefont {Fonseca}, \citenamefont {Friebel},
  \citenamefont {Garolfi}, \citenamefont {Gessner}, \citenamefont {Goddard},
  \citenamefont {Gorgisyan}, \citenamefont {Gorn}, \citenamefont {Granados},
  \citenamefont {Grulke}, \citenamefont {Hartin}, \citenamefont {Helm},
  \citenamefont {Henderson}, \citenamefont {Hüther}, \citenamefont {Ibison},
  \citenamefont {Jolly}, \citenamefont {Keeble}, \citenamefont {Kelisani},
  \citenamefont {Kim}, \citenamefont {Kraus}, \citenamefont {Krupa},
  \citenamefont {Lefevre}, \citenamefont {Li}, \citenamefont {Liu},
  \citenamefont {Lopes}, \citenamefont {Lotov}, \citenamefont {Martyanov},
  \citenamefont {Mazzoni}, \citenamefont {Minakov}, \citenamefont {Molendijk},
  \citenamefont {Moody}, \citenamefont {Moreira}, \citenamefont {Muggli},
  \citenamefont {Panuganti}, \citenamefont {Pardons}, \citenamefont
  {Peña~Asmus}, \citenamefont {Perera}, \citenamefont {Petrenko},
  \citenamefont {Pukhov}, \citenamefont {Rey}, \citenamefont {Sherwood},
  \citenamefont {Silva}, \citenamefont {Sosedkin}, \citenamefont {Tuev},
  \citenamefont {Velotti}, \citenamefont {Verra}, \citenamefont {Verzilov},
  \citenamefont {Vieira}, \citenamefont {Welsch}, \citenamefont {Wendt},
  \citenamefont {Williamson}, \citenamefont {Wing}, \citenamefont {Woolley},
  \citenamefont {Xia},\ and\ \citenamefont {{The AWAKE
  Collaboration}}}]{Gschwendtner_2019_PTRSA_377_20180418}%
  \BibitemOpen
  \bibfield  {author} {\bibinfo {author} {\bibfnamefont {E.}~\bibnamefont
  {Gschwendtner}}, \bibinfo {author} {\bibfnamefont {M.}~\bibnamefont
  {Turner}}, \bibinfo {author} {\bibfnamefont {E.}~\bibnamefont {Adli}},
  \bibinfo {author} {\bibfnamefont {A.}~\bibnamefont {Ahuja}}, \bibinfo
  {author} {\bibfnamefont {O.}~\bibnamefont {Apsimon}}, \bibinfo {author}
  {\bibfnamefont {R.}~\bibnamefont {Apsimon}}, \bibinfo {author} {\bibfnamefont
  {A.-M.}\ \bibnamefont {Bachmann}}, \bibinfo {author} {\bibfnamefont
  {F.}~\bibnamefont {Batsch}}, \bibinfo {author} {\bibfnamefont
  {C.}~\bibnamefont {Bracco}}, \bibinfo {author} {\bibfnamefont
  {F.}~\bibnamefont {Braunmüller}}, \bibinfo {author} {\bibfnamefont
  {S.}~\bibnamefont {Burger}}, \bibinfo {author} {\bibfnamefont
  {G.}~\bibnamefont {Burt}}, \bibinfo {author} {\bibfnamefont {B.}~\bibnamefont
  {Buttenschön}}, \bibinfo {author} {\bibfnamefont {A.}~\bibnamefont
  {Caldwell}}, \bibinfo {author} {\bibfnamefont {J.}~\bibnamefont {Chappell}},
  \bibinfo {author} {\bibfnamefont {E.}~\bibnamefont {Chevallay}}, \bibinfo
  {author} {\bibfnamefont {M.}~\bibnamefont {Chung}}, \bibinfo {author}
  {\bibfnamefont {D.}~\bibnamefont {Cooke}}, \bibinfo {author} {\bibfnamefont
  {H.}~\bibnamefont {Damerau}}, \bibinfo {author} {\bibfnamefont {L.~H.}\
  \bibnamefont {Deubner}}, \bibinfo {author} {\bibfnamefont {A.}~\bibnamefont
  {Dexter}}, \bibinfo {author} {\bibfnamefont {S.}~\bibnamefont {Doebert}},
  \bibinfo {author} {\bibfnamefont {J.}~\bibnamefont {Farmer}}, \bibinfo
  {author} {\bibfnamefont {V.~N.}\ \bibnamefont {Fedosseev}}, \bibinfo {author}
  {\bibfnamefont {R.}~\bibnamefont {Fiorito}}, \bibinfo {author} {\bibfnamefont
  {R.~A.}\ \bibnamefont {Fonseca}}, \bibinfo {author} {\bibfnamefont
  {F.}~\bibnamefont {Friebel}}, \bibinfo {author} {\bibfnamefont
  {L.}~\bibnamefont {Garolfi}}, \bibinfo {author} {\bibfnamefont
  {S.}~\bibnamefont {Gessner}}, \bibinfo {author} {\bibfnamefont
  {B.}~\bibnamefont {Goddard}}, \bibinfo {author} {\bibfnamefont
  {I.}~\bibnamefont {Gorgisyan}}, \bibinfo {author} {\bibfnamefont {A.~A.}\
  \bibnamefont {Gorn}}, \bibinfo {author} {\bibfnamefont {E.}~\bibnamefont
  {Granados}}, \bibinfo {author} {\bibfnamefont {O.}~\bibnamefont {Grulke}},
  \bibinfo {author} {\bibfnamefont {A.}~\bibnamefont {Hartin}}, \bibinfo
  {author} {\bibfnamefont {A.}~\bibnamefont {Helm}}, \bibinfo {author}
  {\bibfnamefont {J.~R.}\ \bibnamefont {Henderson}}, \bibinfo {author}
  {\bibfnamefont {M.}~\bibnamefont {Hüther}}, \bibinfo {author} {\bibfnamefont
  {M.}~\bibnamefont {Ibison}}, \bibinfo {author} {\bibfnamefont
  {S.}~\bibnamefont {Jolly}}, \bibinfo {author} {\bibfnamefont
  {F.}~\bibnamefont {Keeble}}, \bibinfo {author} {\bibfnamefont {M.~D.}\
  \bibnamefont {Kelisani}}, \bibinfo {author} {\bibfnamefont {S.-Y.}\
  \bibnamefont {Kim}}, \bibinfo {author} {\bibfnamefont {F.}~\bibnamefont
  {Kraus}}, \bibinfo {author} {\bibfnamefont {M.}~\bibnamefont {Krupa}},
  \bibinfo {author} {\bibfnamefont {T.}~\bibnamefont {Lefevre}}, \bibinfo
  {author} {\bibfnamefont {Y.}~\bibnamefont {Li}}, \bibinfo {author}
  {\bibfnamefont {S.}~\bibnamefont {Liu}}, \bibinfo {author} {\bibfnamefont
  {N.}~\bibnamefont {Lopes}}, \bibinfo {author} {\bibfnamefont {K.~V.}\
  \bibnamefont {Lotov}}, \bibinfo {author} {\bibfnamefont {M.}~\bibnamefont
  {Martyanov}}, \bibinfo {author} {\bibfnamefont {S.}~\bibnamefont {Mazzoni}},
  \bibinfo {author} {\bibfnamefont {V.~A.}\ \bibnamefont {Minakov}}, \bibinfo
  {author} {\bibfnamefont {J.~C.}\ \bibnamefont {Molendijk}}, \bibinfo {author}
  {\bibfnamefont {J.~T.}\ \bibnamefont {Moody}}, \bibinfo {author}
  {\bibfnamefont {M.}~\bibnamefont {Moreira}}, \bibinfo {author} {\bibfnamefont
  {P.}~\bibnamefont {Muggli}}, \bibinfo {author} {\bibfnamefont
  {H.}~\bibnamefont {Panuganti}}, \bibinfo {author} {\bibfnamefont
  {A.}~\bibnamefont {Pardons}}, \bibinfo {author} {\bibfnamefont
  {F.}~\bibnamefont {Peña~Asmus}}, \bibinfo {author} {\bibfnamefont
  {A.}~\bibnamefont {Perera}}, \bibinfo {author} {\bibfnamefont
  {A.}~\bibnamefont {Petrenko}}, \bibinfo {author} {\bibfnamefont
  {A.}~\bibnamefont {Pukhov}}, \bibinfo {author} {\bibfnamefont
  {S.}~\bibnamefont {Rey}}, \bibinfo {author} {\bibfnamefont {P.}~\bibnamefont
  {Sherwood}}, \bibinfo {author} {\bibfnamefont {L.~O.}\ \bibnamefont {Silva}},
  \bibinfo {author} {\bibfnamefont {A.~P.}\ \bibnamefont {Sosedkin}}, \bibinfo
  {author} {\bibfnamefont {P.~V.}\ \bibnamefont {Tuev}}, \bibinfo {author}
  {\bibfnamefont {F.}~\bibnamefont {Velotti}}, \bibinfo {author} {\bibfnamefont
  {L.}~\bibnamefont {Verra}}, \bibinfo {author} {\bibfnamefont {V.~A.}\
  \bibnamefont {Verzilov}}, \bibinfo {author} {\bibfnamefont {J.}~\bibnamefont
  {Vieira}}, \bibinfo {author} {\bibfnamefont {C.~P.}\ \bibnamefont {Welsch}},
  \bibinfo {author} {\bibfnamefont {M.}~\bibnamefont {Wendt}}, \bibinfo
  {author} {\bibfnamefont {B.}~\bibnamefont {Williamson}}, \bibinfo {author}
  {\bibfnamefont {M.}~\bibnamefont {Wing}}, \bibinfo {author} {\bibfnamefont
  {B.}~\bibnamefont {Woolley}}, \bibinfo {author} {\bibfnamefont
  {G.}~\bibnamefont {Xia}}, \ and\ \bibinfo {author} {\bibnamefont {{The AWAKE
  Collaboration}}},\ }\href {\doibase 10.1098/rsta.2018.0418} {\bibfield
  {journal} {\bibinfo  {journal} {Phil. Trans. R. Soc. A}\ }\textbf {\bibinfo
  {volume} {377}},\ \bibinfo {pages} {20180418} (\bibinfo {year}
  {2019})}\BibitemShut {NoStop}%
\bibitem [{\citenamefont {Adli}\ and\ \citenamefont
  {Muggli}(2016)}]{Adli_2016_RAST_09_85}%
  \BibitemOpen
  \bibfield  {author} {\bibinfo {author} {\bibfnamefont {E.}~\bibnamefont
  {Adli}}\ and\ \bibinfo {author} {\bibfnamefont {P.}~\bibnamefont {Muggli}},\
  }\href {\doibase 10.1142/S1793626816300048} {\bibfield  {journal} {\bibinfo
  {journal} {Rev. Accel. Sci. Tech.}\ }\textbf {\bibinfo {volume} {09}},\
  \bibinfo {pages} {85} (\bibinfo {year} {2016})}\BibitemShut {NoStop}%
\bibitem [{\citenamefont {Birdsall}\ and\ \citenamefont
  {Langdon}(2004)}]{Birdsall_2004}%
  \BibitemOpen
  \bibfield  {author} {\bibinfo {author} {\bibfnamefont {C.~K.}\ \bibnamefont
  {Birdsall}}\ and\ \bibinfo {author} {\bibfnamefont {A.~B.}\ \bibnamefont
  {Langdon}},\ }\href@noop {} {\emph {\bibinfo {title} {Plasma physics via
  computer simulation}}}\ (\bibinfo  {publisher} {CRC press},\ \bibinfo {year}
  {2004})\BibitemShut {NoStop}%
\bibitem [{\citenamefont {Pukhov}(2016)}]{Pukhov2016PIC}%
  \BibitemOpen
  \bibfield  {author} {\bibinfo {author} {\bibfnamefont {A.}~\bibnamefont
  {Pukhov}},\ }\href {\doibase 10.5170/CERN-2016-001.181} {\bibfield  {journal}
  {\bibinfo  {journal} {CERN Yellow Rep.}\ }\textbf {\bibinfo {volume} {1}},\
  \bibinfo {pages} {181} (\bibinfo {year} {2016})}\BibitemShut {NoStop}%
\bibitem [{\citenamefont {Gorbunov}\ and\ \citenamefont
  {Kirsanov}(1987)}]{Gorbunov_1987_JETP_66_290}%
  \BibitemOpen
  \bibfield  {author} {\bibinfo {author} {\bibfnamefont {L.~M.}\ \bibnamefont
  {Gorbunov}}\ and\ \bibinfo {author} {\bibfnamefont {V.~I.}\ \bibnamefont
  {Kirsanov}},\ }\href@noop {} {\bibfield  {journal} {\bibinfo  {journal} {Sov.
  Phys. JETP}\ }\textbf {\bibinfo {volume} {66}},\ \bibinfo {pages} {290}
  (\bibinfo {year} {1987})}\BibitemShut {NoStop}%
\bibitem [{\citenamefont {Chen}\ \emph {et~al.}(1987)\citenamefont {Chen},
  \citenamefont {Su}, \citenamefont {Katsouleas}, \citenamefont {Wilks},\ and\
  \citenamefont {Dawson}}]{Chen_1987_IEEETPS_15_218}%
  \BibitemOpen
  \bibfield  {author} {\bibinfo {author} {\bibfnamefont {P.}~\bibnamefont
  {Chen}}, \bibinfo {author} {\bibfnamefont {J.~J.}\ \bibnamefont {Su}},
  \bibinfo {author} {\bibfnamefont {T.}~\bibnamefont {Katsouleas}}, \bibinfo
  {author} {\bibfnamefont {S.}~\bibnamefont {Wilks}}, \ and\ \bibinfo {author}
  {\bibfnamefont {J.~M.}\ \bibnamefont {Dawson}},\ }\href {\doibase
  10.1109/TPS.1987.4316688} {\bibfield  {journal} {\bibinfo  {journal} {IEEE
  Trans. Plasma Sci.}\ }\textbf {\bibinfo {volume} {15}},\ \bibinfo {pages}
  {218} (\bibinfo {year} {1987})}\BibitemShut {NoStop}%
\bibitem [{\citenamefont {Lotov}(2004)}]{Lotov_2004_PRL_69_046405}%
  \BibitemOpen
  \bibfield  {author} {\bibinfo {author} {\bibfnamefont {K.~V.}\ \bibnamefont
  {Lotov}},\ }\href {\doibase 10.1103/PhysRevE.69.046405} {\bibfield  {journal}
  {\bibinfo  {journal} {Phys. Rev. E}\ }\textbf {\bibinfo {volume} {69}},\
  \bibinfo {pages} {046405} (\bibinfo {year} {2004})}\BibitemShut {NoStop}%
\bibitem [{\citenamefont {Kostyukov}, \citenamefont {Pukhov},\ and\
  \citenamefont {Kiselev}(2004)}]{Kostyukov_2004_PoP_11_115256}%
  \BibitemOpen
  \bibfield  {author} {\bibinfo {author} {\bibfnamefont {I.}~\bibnamefont
  {Kostyukov}}, \bibinfo {author} {\bibfnamefont {A.}~\bibnamefont {Pukhov}}, \
  and\ \bibinfo {author} {\bibfnamefont {S.}~\bibnamefont {Kiselev}},\ }\href
  {\doibase 10.1063/1.1799371} {\bibfield  {journal} {\bibinfo  {journal}
  {Phys. Plasmas}\ }\textbf {\bibinfo {volume} {11}},\ \bibinfo {pages} {5256}
  (\bibinfo {year} {2004})}\BibitemShut {NoStop}%
\bibitem [{\citenamefont {Lu}\ \emph {et~al.}(2006)\citenamefont {Lu},
  \citenamefont {Huang}, \citenamefont {Zhou}, \citenamefont {Mori},\ and\
  \citenamefont {Katsouleas}}]{Lu_2006_PRL_96_165002}%
  \BibitemOpen
  \bibfield  {author} {\bibinfo {author} {\bibfnamefont {W.}~\bibnamefont
  {Lu}}, \bibinfo {author} {\bibfnamefont {C.}~\bibnamefont {Huang}}, \bibinfo
  {author} {\bibfnamefont {M.}~\bibnamefont {Zhou}}, \bibinfo {author}
  {\bibfnamefont {W.~B.}\ \bibnamefont {Mori}}, \ and\ \bibinfo {author}
  {\bibfnamefont {T.}~\bibnamefont {Katsouleas}},\ }\href {\doibase
  10.1103/PhysRevLett.96.165002} {\bibfield  {journal} {\bibinfo  {journal}
  {Phys. Rev. Lett.}\ }\textbf {\bibinfo {volume} {96}},\ \bibinfo {pages}
  {165002} (\bibinfo {year} {2006})}\BibitemShut {NoStop}%
\bibitem [{\citenamefont {Golovanov}\ \emph
  {et~al.}(2016{\natexlab{a}})\citenamefont {Golovanov}, \citenamefont
  {Kostyukov}, \citenamefont {Pukhov},\ and\ \citenamefont
  {Thomas}}]{Golovanov_2016_QE_46_295}%
  \BibitemOpen
  \bibfield  {author} {\bibinfo {author} {\bibfnamefont {A.~A.}\ \bibnamefont
  {Golovanov}}, \bibinfo {author} {\bibfnamefont {I.~{\relax Yu}.}\
  \bibnamefont {Kostyukov}}, \bibinfo {author} {\bibfnamefont {A.~M.}\
  \bibnamefont {Pukhov}}, \ and\ \bibinfo {author} {\bibfnamefont
  {J.}~\bibnamefont {Thomas}},\ }\href {\doibase 10.1070/QEL16040} {\bibfield
  {journal} {\bibinfo  {journal} {Quantum Electron.}\ }\textbf {\bibinfo
  {volume} {46}},\ \bibinfo {pages} {295} (\bibinfo {year}
  {2016}{\natexlab{a}})}\BibitemShut {NoStop}%
\bibitem [{\citenamefont {Tzoufras}\ \emph {et~al.}(2009)\citenamefont
  {Tzoufras}, \citenamefont {Lu}, \citenamefont {Tsung}, \citenamefont {Huang},
  \citenamefont {Mori}, \citenamefont {Katsouleas}, \citenamefont {Vieira},
  \citenamefont {Fonseca},\ and\ \citenamefont
  {Silva}}]{Tzoufras_2009_PoP_16_056705}%
  \BibitemOpen
  \bibfield  {author} {\bibinfo {author} {\bibfnamefont {M.}~\bibnamefont
  {Tzoufras}}, \bibinfo {author} {\bibfnamefont {W.}~\bibnamefont {Lu}},
  \bibinfo {author} {\bibfnamefont {F.~S.}\ \bibnamefont {Tsung}}, \bibinfo
  {author} {\bibfnamefont {C.}~\bibnamefont {Huang}}, \bibinfo {author}
  {\bibfnamefont {W.~B.}\ \bibnamefont {Mori}}, \bibinfo {author}
  {\bibfnamefont {T.}~\bibnamefont {Katsouleas}}, \bibinfo {author}
  {\bibfnamefont {J.}~\bibnamefont {Vieira}}, \bibinfo {author} {\bibfnamefont
  {R.~A.}\ \bibnamefont {Fonseca}}, \ and\ \bibinfo {author} {\bibfnamefont
  {L.~O.}\ \bibnamefont {Silva}},\ }\href {\doibase 10.1063/1.3118628}
  {\bibfield  {journal} {\bibinfo  {journal} {Phys. Plasmas}\ }\textbf
  {\bibinfo {volume} {16}},\ \bibinfo {pages} {056705} (\bibinfo {year}
  {2009})}\BibitemShut {NoStop}%
\bibitem [{\citenamefont {Thomas}\ \emph {et~al.}(2016)\citenamefont {Thomas},
  \citenamefont {Kostyukov}, \citenamefont {Pronold}, \citenamefont
  {Golovanov},\ and\ \citenamefont {Pukhov}}]{Thomas_2016_PoP_23_053108}%
  \BibitemOpen
  \bibfield  {author} {\bibinfo {author} {\bibfnamefont {J.}~\bibnamefont
  {Thomas}}, \bibinfo {author} {\bibfnamefont {I.~{\relax Yu}.}\ \bibnamefont
  {Kostyukov}}, \bibinfo {author} {\bibfnamefont {J.}~\bibnamefont {Pronold}},
  \bibinfo {author} {\bibfnamefont {A.}~\bibnamefont {Golovanov}}, \ and\
  \bibinfo {author} {\bibfnamefont {A.}~\bibnamefont {Pukhov}},\ }\href
  {\doibase 10.1063/1.4948712} {\bibfield  {journal} {\bibinfo  {journal}
  {Phys. Plasmas}\ }\textbf {\bibinfo {volume} {23}},\ \bibinfo {pages}
  {053108} (\bibinfo {year} {2016})}\BibitemShut {NoStop}%
\bibitem [{\citenamefont {Golovanov}\ \emph
  {et~al.}(2016{\natexlab{b}})\citenamefont {Golovanov}, \citenamefont
  {Kostyukov}, \citenamefont {Thomas},\ and\ \citenamefont
  {Pukhov}}]{Golovanov_2016_PoP_23_093114}%
  \BibitemOpen
  \bibfield  {author} {\bibinfo {author} {\bibfnamefont {A.~A.}\ \bibnamefont
  {Golovanov}}, \bibinfo {author} {\bibfnamefont {I.~{\relax Yu}.}\
  \bibnamefont {Kostyukov}}, \bibinfo {author} {\bibfnamefont {J.}~\bibnamefont
  {Thomas}}, \ and\ \bibinfo {author} {\bibfnamefont {A.}~\bibnamefont
  {Pukhov}},\ }\href {\doibase 10.1063/1.4962565} {\bibfield  {journal}
  {\bibinfo  {journal} {Phys. Plasmas}\ }\textbf {\bibinfo {volume} {23}},\
  \bibinfo {pages} {093114} (\bibinfo {year} {2016}{\natexlab{b}})}\BibitemShut
  {NoStop}%
\bibitem [{\citenamefont {Sprangle}, \citenamefont {Esarey},\ and\
  \citenamefont {Ting}(1990)}]{Sprangle_1990_PRL_64_2011}%
  \BibitemOpen
  \bibfield  {author} {\bibinfo {author} {\bibfnamefont {P.}~\bibnamefont
  {Sprangle}}, \bibinfo {author} {\bibfnamefont {E.}~\bibnamefont {Esarey}}, \
  and\ \bibinfo {author} {\bibfnamefont {A.}~\bibnamefont {Ting}},\ }\href
  {\doibase 10.1103/PhysRevLett.64.2011} {\bibfield  {journal} {\bibinfo
  {journal} {Phys. Rev. Lett.}\ }\textbf {\bibinfo {volume} {64}},\ \bibinfo
  {pages} {2011} (\bibinfo {year} {1990})}\BibitemShut {NoStop}%
\bibitem [{\citenamefont {Yi}\ \emph {et~al.}(2013)\citenamefont {Yi},
  \citenamefont {Khudik}, \citenamefont {Siemon},\ and\ \citenamefont
  {Shvets}}]{Yi_2013_PoP_20_013108}%
  \BibitemOpen
  \bibfield  {author} {\bibinfo {author} {\bibfnamefont {S.~A.}\ \bibnamefont
  {Yi}}, \bibinfo {author} {\bibfnamefont {V.}~\bibnamefont {Khudik}}, \bibinfo
  {author} {\bibfnamefont {C.}~\bibnamefont {Siemon}}, \ and\ \bibinfo {author}
  {\bibfnamefont {G.}~\bibnamefont {Shvets}},\ }\href {\doibase
  10.1063/1.4775774} {\bibfield  {journal} {\bibinfo  {journal} {Phys.
  Plasmas}\ }\textbf {\bibinfo {volume} {20}},\ \bibinfo {pages} {013108}
  (\bibinfo {year} {2013})}\BibitemShut {NoStop}%
\bibitem [{\citenamefont {Golovanov}\ \emph {et~al.}(2017)\citenamefont
  {Golovanov}, \citenamefont {Kostyukov}, \citenamefont {Thomas},\ and\
  \citenamefont {Pukhov}}]{Golovanov_2017_PoP_24_103104}%
  \BibitemOpen
  \bibfield  {author} {\bibinfo {author} {\bibfnamefont {A.~A.}\ \bibnamefont
  {Golovanov}}, \bibinfo {author} {\bibfnamefont {I.~{\relax Yu}.}\
  \bibnamefont {Kostyukov}}, \bibinfo {author} {\bibfnamefont {J.}~\bibnamefont
  {Thomas}}, \ and\ \bibinfo {author} {\bibfnamefont {A.}~\bibnamefont
  {Pukhov}},\ }\href {\doibase 10.1063/1.4996856} {\bibfield  {journal}
  {\bibinfo  {journal} {Phys. Plasmas}\ }\textbf {\bibinfo {volume} {24}},\
  \bibinfo {pages} {103104} (\bibinfo {year} {2017})}\BibitemShut {NoStop}%
\bibitem [{\citenamefont {Litvak}(1970)}]{Litvak_1970_JETP_30_166}%
  \BibitemOpen
  \bibfield  {author} {\bibinfo {author} {\bibfnamefont {A.}~\bibnamefont
  {Litvak}},\ }\href@noop {} {\bibfield  {journal} {\bibinfo  {journal} {Sov.
  Phys. JETP}\ }\textbf {\bibinfo {volume} {30}},\ \bibinfo {pages} {166}
  (\bibinfo {year} {1970})}\BibitemShut {NoStop}%
\bibitem [{\citenamefont {{Sprangle}}, \citenamefont {{Tang}},\ and\
  \citenamefont {{Esarey}}(1987)}]{Sprangle_1987_IEEETPS_15_145}%
  \BibitemOpen
  \bibfield  {author} {\bibinfo {author} {\bibfnamefont {P.}~\bibnamefont
  {{Sprangle}}}, \bibinfo {author} {\bibfnamefont {C.}~\bibnamefont {{Tang}}},
  \ and\ \bibinfo {author} {\bibfnamefont {E.}~\bibnamefont {{Esarey}}},\
  }\href {\doibase 10.1109/TPS.1987.4316677} {\bibfield  {journal} {\bibinfo
  {journal} {IEEE Trans. Plasma Sci.}\ }\textbf {\bibinfo {volume} {15}},\
  \bibinfo {pages} {145} (\bibinfo {year} {1987})}\BibitemShut {NoStop}%
\bibitem [{\citenamefont {Reichwein}\ \emph {et~al.}(2020)\citenamefont
  {Reichwein}, \citenamefont {Thomas}, \citenamefont {Golovanov}, \citenamefont
  {Kostyukov},\ and\ \citenamefont {Pukhov}}]{Reichwein_2020_PPCF_62_115017}%
  \BibitemOpen
  \bibfield  {author} {\bibinfo {author} {\bibfnamefont {L.}~\bibnamefont
  {Reichwein}}, \bibinfo {author} {\bibfnamefont {J.}~\bibnamefont {Thomas}},
  \bibinfo {author} {\bibfnamefont {A.}~\bibnamefont {Golovanov}}, \bibinfo
  {author} {\bibfnamefont {I.}~\bibnamefont {Kostyukov}}, \ and\ \bibinfo
  {author} {\bibfnamefont {A.}~\bibnamefont {Pukhov}},\ }\href {\doibase
  10.1088/1361-6587/abb618} {\bibfield  {journal} {\bibinfo  {journal} {Plasma
  Phys. Control. Fusion}\ }\textbf {\bibinfo {volume} {62}},\ \bibinfo {pages}
  {115017} (\bibinfo {year} {2020})}\BibitemShut {NoStop}%
\bibitem [{\citenamefont {Derouillat}\ \emph {et~al.}(2018)\citenamefont
  {Derouillat}, \citenamefont {Beck}, \citenamefont {Pérez}, \citenamefont
  {Vinci}, \citenamefont {Chiaramello}, \citenamefont {Grassi}, \citenamefont
  {Flé}, \citenamefont {Bouchard}, \citenamefont {Plotnikov}, \citenamefont
  {Aunai}, \citenamefont {Dargent}, \citenamefont {Riconda},\ and\
  \citenamefont {Grech}}]{Derouillat_2018_CPC_222_351}%
  \BibitemOpen
  \bibfield  {author} {\bibinfo {author} {\bibfnamefont {J.}~\bibnamefont
  {Derouillat}}, \bibinfo {author} {\bibfnamefont {A.}~\bibnamefont {Beck}},
  \bibinfo {author} {\bibfnamefont {F.}~\bibnamefont {Pérez}}, \bibinfo
  {author} {\bibfnamefont {T.}~\bibnamefont {Vinci}}, \bibinfo {author}
  {\bibfnamefont {M.}~\bibnamefont {Chiaramello}}, \bibinfo {author}
  {\bibfnamefont {A.}~\bibnamefont {Grassi}}, \bibinfo {author} {\bibfnamefont
  {M.}~\bibnamefont {Flé}}, \bibinfo {author} {\bibfnamefont {G.}~\bibnamefont
  {Bouchard}}, \bibinfo {author} {\bibfnamefont {I.}~\bibnamefont {Plotnikov}},
  \bibinfo {author} {\bibfnamefont {N.}~\bibnamefont {Aunai}}, \bibinfo
  {author} {\bibfnamefont {J.}~\bibnamefont {Dargent}}, \bibinfo {author}
  {\bibfnamefont {C.}~\bibnamefont {Riconda}}, \ and\ \bibinfo {author}
  {\bibfnamefont {M.}~\bibnamefont {Grech}},\ }\href {\doibase
  10.1016/j.cpc.2017.09.024} {\bibfield  {journal} {\bibinfo  {journal} {Comp.
  Phys. Commun.}\ }\textbf {\bibinfo {volume} {222}},\ \bibinfo {pages} {351}
  (\bibinfo {year} {2018})}\BibitemShut {NoStop}%
\bibitem [{Smi()}]{Smilei}%
  \BibitemOpen
  \href@noop {} {}\bibinfo {howpublished} {Smilei,
  \url{http://www.maisondelasimulation.fr/smilei/}}\BibitemShut {NoStop}%
\end{thebibliography}%

\end{document}